\documentclass[a4paper, 12pt]{article}
\usepackage{amssymb}
\usepackage{amsthm}
\usepackage{bm}
\usepackage{caption}
\usepackage{here}
\usepackage{mathtools}
\usepackage{multirow}
\usepackage{url}
\usepackage{color}

\usepackage{hyperref}
\hypersetup{
	colorlinks=true,
	citecolor=blue,
	linkcolor=blue,
	urlcolor=blue
}
\theoremstyle{thmstyleone}
\newtheorem{theorem}{Theorem}
\newtheorem{proposition}{Proposition}
\newtheorem{corollary}{Corollary}
\newtheorem{lemma}{Lemma}

\theoremstyle{remark}
\newtheorem{remark}{Remark}
\newtheorem{step}{Step}

\begin{document}
	\title{Mixed effects models for extreme value index regression}
	\author{{\sc{Koki Momoki}}$^{1}$\thanks{E-mail: k3499390@kadai.jp} and {\sc{Takuma Yoshida}}$^{2}$\thanks{E-mail: yoshida@sci.kagoshima-u.ac.jp}\\\\
		$^{1,2}${\it{Graduate School of Science and Engineering, Kagoshima University}}\\
		{\it{1-21-40 Korimoto, Kagoshima, Kagoshima, 890-8580, Japan}}}
	
	\date{}
	
	\maketitle
	
	\abstract{Extreme value theory (EVT) provides an elegant mathematical tool for the statistical analysis of rare events. When data are collected from multiple population subgroups, because some subgroups may have less data available for extreme value analysis, a scientific interest of many researchers would be to improve the estimates obtained directly from each subgroup. To achieve this, we incorporate the mixed effects model (MEM) into the regression technique in EVT. In small area estimation, the MEM has attracted considerable attention as a primary tool for producing reliable estimates for subgroups with small sample sizes, i.e., ``small areas.'' The key idea of MEM is to incorporate information from all subgroups into a single model and to borrow strength from all subgroups to improve estimates for each subgroup. Using this property, in extreme value analysis, the MEM may contribute to reducing the bias and variance of the direct estimates from each subgroup. This prompts us to evaluate the effectiveness of the MEM for EVT through theoretical studies and numerical experiments, including its application to the risk assessment of a number of stocks in the cryptocurrency market.}
	
	\bigskip
	
	\noindent
	{\it{Keywords: Extreme value index regression; Extreme value theory; Mixed effects model; Pareto-type distribution; Risk assessment}}
	
	\section{Introduction}\label{Sec1}
	
	Statistical analysis of rare events is crucial for risk assessment in various fields, including meteorology, environment, seismology, finance, economics and insurance. Extreme value theory (EVT) provides an elegant mathematical tool for the analysis of rare events.
	
	In the framework of univariate EVT, the generalized extreme value distribution (see, Fisher and Tippett 1928; Gumbel 1958) and generalized Pareto distribution (GPD, see, Davison and Smith 1990) are standard models for fitting extreme value data. In addition, the class of GPDs with unbounded tails is called the Pareto-type distribution, which has been recognized as an important model for analyzing heavy-tailed data (see, Beirlant et al. 2004). As an important direction of development in EVT, these models have been extended to regression models to incorporate covariate information into extreme value analysis (see, Davison and Smith 1990; Beirlant and Goegebeur 2003; Beirlant et al. 2004; Wang and Tsai 2009). However, few regression techniques have been developed for unit-level data in extreme value analysis.
	
	The data category of interest is denoted by $\{(Y_{ij}, {\bm{X}}_{ij}),\ i=1, 2, \ldots, n_j,\ j=1, 2, \ldots, J\}$, where $J$ is the number of population subgroups, $n_j$ is the sample size from the $j$th population subgroup, $Y_{ij}$ is the response of interest, and ${\bm{X}}_{ij}$ is the vector of relevant covariates. In much of the literature on small area estimation (SAE), this data category is referred to as unit-level data, and each subgroup is then called an ``area'' (see, Rao and Molina 2015; Sugasawa and Kubokawa 2020; Molina, Corral, and Nguyen 2022). In particular, an area with a small sample size is the so-called ``small area,'' which is described in detail by Jiang (2017). Examples of an area include a geographic region such as a state, county or municipality, or a demographic group such as a specific age-sex-race group. From the definition on p. 173 of Rao and Molina (2015), unit-level data means that the number of areas $J$ is known, and each observation $(Y_{ij}, {\bm{X}}_{ij})$ is explicitly assigned to one of the areas. This study aims to develop the regression technique of extreme value analysis for unit-level data.
	
	Our purpose is not to pool data from multiple areas into a single area by clustering, nor to create a new set of areas with heterogeneous characteristics (see, Bottolo et al. 2003; Rohrbeck and Tawn 2021; de Carvalho, Huser, and Rubio 2023; Dupuis, Engelke, and Trapin 2023 and references therein). Unlike these approaches, our goal is to enhance the accuracy of extreme value analysis by using information from all areas simultaneously, instead of building models by area. However, if the heterogeneity between all areas is modeled as parameters, the number of parameters depends on the number of areas $J$. Accordingly, for large $J$, the fully parametric model can lead to a large bias in the parameter estimates (see, Section 4 of Ruppert, Wand, and Carroll 2003; Brostr\"{o}m and Holmberg 2011). Thus, we want to develop a model for extreme value analysis that does not require many parameters for unit-level data. To this end, we incorporate the mixed effects model (MEM) into EVT. The MEM has been described in Jiang (2007), Wu (2009), Jiang (2017), and references therein. This model captures the heterogeneity between areas as a latent structure rather than as parameters. In SAE (see, Torabi 2019; Sugasawa and Kubokawa 2020), the efficiency of MEM is well known for its so-called ``borrowing of strength'' property (see, Dempster, Rubin, and Tsutakawa 1981). Dempster, Rubin, and Tsutakawa (1981) described the ``borrowing of strength'' as follows:
	\begin{quote}
		Using concepts of variability between and within units, one can improve estimates based on the data from a single unit by the appropriate use of data from the remaining units. (Dempster, Rubin, and Tsutakawa 1981, p. 341)
	\end{quote}
	In other words, the ``borrowing of strength'' indicates that direct estimates based only on area-specific data are improved by using information from other areas (see, Section 1 of Diallo and Rao 2018). For small areas, the ``borrowing of strength'' would be particularly helpful because the accuracy of their direct estimates is not sufficiently guaranteed (see, Molina and Rao 2010; Diallo and Rao 2018). In extreme value analysis, we use only data with small or large values; hence, the effective sample size tends to be small for some areas. Thus, the MEM would be crucial to obtain more efficient estimators for extreme value analysis. However, to the best of our knowledge, there are no results for combining the MEM with EVT. Therefore, it would be important to show that the ``borrowing of strength'' is also valid for extreme value analysis. We will reveal such considerations theoretically and numerically.
	
	In this study, we incorporate the MEM into the extreme value index (EVI) of the Pareto-type distribution. For this model, we first pick the extreme value data for each area using the peak-over-threshold method. Then, the regression parameters are estimated by the maximum likelihood method. In addition, the random components of the MEM, called random effects, which correspond to the latent area-wise differences in the regression coefficients, are predicted by the conditional mode. We investigate the asymptotic normality of the proposed estimator (see, Section \ref{Sec3.2}). From this asymptotic normality, we find that the variance of the estimator improves as the number of areas $J$ increases. In other words, the proposed estimator is generally stable even when the effective sample size is small for certain areas. Owing to this property, the proposed estimator can reduce the severe bias of Pareto-type modeling by setting reasonably higher thresholds while achieving its stable behavior. Furthermore, we show numerically through the Monte Carlo simulation study that our estimates for each area, obtained by combining the proposed estimator and predictor, not only have significantly smaller variances than the direct estimates, but are also less biased (see, Section 1.3 of our supplementary material). Surprisingly, in the context of EVT, the ``borrowing of strength'' of the MEM contributes to reducing both bias and variance.
	
	As an application, we analyze the returns of 413 cryptocurrencies. Since their risks as assets may vary by stock, an accurate risk assessment for each stock would be highly desirable. Along with some analysis, we will demonstrate the effectiveness of the MEM in analyzing the risks of many cryptocurrency stocks. We note that considering many stocks is equivalent to the mathematical situation that the number of areas $J$ is large. Thus, the theoretical study of the proposed estimator will be established under $J\to\infty$.
	
	The remainder of this article is organized as follows. Section \ref{Sec2} proposes the regression model using the MEM for the Pareto-type distribution. Section \ref{Sec3} examines its asymptotic properties. As a real data example, Section \ref{Sec4} analyzes a cryptocurrency dataset. Section \ref{Sec5} summarizes this study and discusses future research. The simulation studies for the proposed model and technical details of the asymptotic theory in Section \ref{Sec3} are described in our supplementary material.
	
	\section{Model and method}\label{Sec2}
	
	Let $\mathbb{R}^+$ be defined as the set of all positive real numbers.  Throughout this article, we consider the unit-level data
	\begin{equation}
		\left\{(Y_{ij}, {\bm{X}}_{ij})\in\mathbb{R}^+\times\mathbb{R}^p,\ i=1, 2, \ldots, n_j,\ j=1, 2, \ldots, J\right\},\label{Eq2.0.1}
	\end{equation}
	where $J$ is the number of areas, $n_j$ is the sample size from the $j$th area, $Y_{ij}$ is the continuous random variable corresponding to the response of interest, and ${\bm{X}}_{ij}$ is the random vector representing the associated predictors. Here, $(Y_{ij}, {\bm{X}}_{ij})$ is regarded as the observation for the $i$th unit in the $j$th area. We denote the index sets by $\mathcal{N}_j\coloneqq\{1, 2, \ldots, n_j\}$ and $\mathcal{J}\coloneqq\{1, 2, \ldots, J\}$. In the following Sections \ref{Sec2.1}-\ref{Sec2.4}, we describe the proposed MEM and associated estimation and prediction methods.
	
	\subsection{Mixed effects model under Pareto-type distribution}\label{Sec2.1}
	
	Let ${\bm{X}}_{ij},\ i\in\mathcal{N}_j,\ j\in\mathcal{J}$ be an independent and identically distributed (i.i.d.) random sample from some distribution. Subsequently, we assume that for each $i\in\mathcal{N}_j$ and $j\in\mathcal{J}$, the response $Y_{ij}$ is conditionally independently obtained from a certain conditional distribution $F_j(y\mid {\bm{x}})=P(Y_{ij}\leq y\mid {\bm{X}}_{ij}={\bm{x}})$ for the given ${\bm{X}}_{ij}$, where $F_j$ is determined for each $j\in\mathcal{J}$. In this study, we are interested in the right tail behavior of each $F_j$. Here, the right tail of each $F_j$ is modeled by the Pareto-type distribution as
	\begin{equation}
		F_j(y\mid {\bm{x}})=1-y^{-1/\gamma_j({\bm{x}})}\mathcal{L}_j(y; {\bm{x}}),\quad j\in\mathcal{J},\label{Eq2.1.1}
	\end{equation}
	where $\gamma_j({\bm{x}})>0$ is a positive function called EVI, and $\mathcal{L}_j(y; {\bm{x}})$ is a conditional slowly varying function with respect to $y$ given ${\bm{x}}$, i.e., for any ${\bm{x}}$ and $s>0$, $\mathcal{L}_j(ys; {\bm{x}})/\mathcal{L}_j(y; {\bm{x}})\to1$ as $y\to\infty$. The EVI function $\gamma_j({\bm{x}})$, which determines the heaviness of the right tail of $F_j$, is assumed here to be the classical linear model formulated as follows:
	\begin{equation}
		\gamma_j({\bm{x}})=\exp\left[\left({\bm{\theta}}_{j{\rm{A}}}^0\right)^\top{\bm{x}}_{\rm{A}}+\left({\bm{\theta}}_{\rm{B}}^0\right)^\top{\bm{x}}_{\rm{B}}\right],\quad j\in\mathcal{J}, \label{Eq2.1.2}
	\end{equation}
	where ${\bm{x}}\coloneqq({\bm{x}}_{\rm{A}}^\top, {\bm{x}}_{\rm{B}}^\top)^\top\in\mathbb{R}^{p_{\rm{A}}}\times\mathbb{R}^{p_{\rm{B}}}$, and ${\bm{\theta}}_{j{\rm{A}}}^0\in\mathbb{R}^{p_{\rm{A}}}$ and ${\bm{\theta}}_{\rm{B}}^0\in\mathbb{R}^{p_{\rm{B}}}$ are regression coefficient vectors. Note that ${\bm{\theta}}_{j{\rm{A}}}^0$ is different between areas, whereas ${\bm{\theta}}_{{\rm{B}}}^0$ is common to all areas. When $J=1$, the above model is reduced to that of Wang and Tsai (2009). The purpose of the model (\ref{Eq2.1.1}) with (\ref{Eq2.1.2}) is to estimate the parameter vectors ${\bm{\theta}}_{1{\rm{A}}}^0, {\bm{\theta}}_{2{\rm{A}}}^0, \ldots, {\bm{\theta}}_{J{\rm{A}}}^0$ and ${\bm{\theta}}_{\rm{B}}^0$. However, this model has $(J\times p_{\rm{A}}+p_{\rm{B}})$ parameters; hence, if $J$ is large, the associated estimators may be severely biased (see, Section 4 of Ruppert, Wand, and Carroll 2003; Brostr\"{o}m and Holmberg 2011). To overcome this bias problem, we use the MEM instead of the fully parametric model (\ref{Eq2.1.1}) with (\ref{Eq2.1.2}).
	
	For $p_{\rm{A}}\leq p$, we introduce the random effects ${\bm{U}}_j\in\mathbb{R}^{p_{\rm{A}}},\ j\in\mathcal{J}$ such that
	\begin{equation}
		{\bm{U}}_1, {\bm{U}}_2, \ldots, {\bm{U}}_J\overset{{\rm{i.i.d.}}}{\sim} N({\bm{0}}, {\bm{\Sigma}}_0),\label{Eq2.1.3}
	\end{equation}
	where $N({\bm{0}}, {\bm{\Sigma}}_0)$ refers to the multivariate normal distribution with zero mean vector and unknown covariance matrix ${\bm{\Sigma}}_0$. The MEM uses these random effects to express the differences in $F_j$ between areas as a latent structure. Let $F(y\mid {\bm{u}}_j, {\bm{x}})\coloneqq P(Y_{ij}\leq y\mid {\bm{U}}_j={\bm{u}}_j, {\bm{X}}_{ij}={\bm{x}})$ be the conditional distribution function of $Y_{ij}$ given ${\bm{U}}_j={\bm{u}}_j$ and ${\bm{X}}_{ij}={\bm{x}}$. In this expression, the information about the differences in $F_j$ between areas is assigned to $F$ by ${\bm{u}}_j$. Note that the random effects ${\bm{U}}_j,\ j\in\mathcal{J}$ are not observed as data.
	
	As an alternative model to (\ref{Eq2.1.1}) with (\ref{Eq2.1.2}), the Pareto-type distribution using the MEM is defined as follows:
	\begin{equation}
		F(y\mid {\bm{u}}_j, {\bm{x}})=1-y^{-1/\gamma({\bm{u}}_j, {\bm{x}})}\mathcal{L}(y; {\bm{u}}_j, {\bm{x}}),\quad j\in\mathcal{J},\label{Eq2.1.4}
	\end{equation}
	where $\mathcal{L}(y; {\bm{u}}_j, {\bm{x}})$ conditional on ${\bm{U}}_j={\bm{u}}_j$ and ${\bm{X}}_{ij}={\bm{x}}$ is a slowly varying function with respect to $y$. Then, as an extension of (\ref{Eq2.1.2}) to the MEM, the EVI function $\gamma({\bm{u}}_j, {\bm{x}})$ is assumed to be
	\begin{equation}
		\gamma({\bm{u}}_j, {\bm{x}})=\exp\left[\left({\bm{\theta}}_{\rm{A}}^0+{\bm{u}}_j\right)^\top{\bm{x}}_{\rm{A}}+\left({\bm{\theta}}_{\rm{B}}^0\right)^\top{\bm{x}}_{\rm{B}}\right],\quad j\in\mathcal{J}.\label{Eq2.1.5}
	\end{equation}
	Compared to (\ref{Eq2.1.2}), the area-wise differences in the slopes of log-EVI with respect to ${\bm{X}}_{{\rm{A}}ij}$ are represented by ${\bm{u}}_j,\ j\in\mathcal{J}$ as a latent structure. Thus, the total number of parameters in the model (\ref{Eq2.1.4}) using (\ref{Eq2.1.5}) is $p+p_{\rm{A}}(p_{\rm{A}}+1)/2$, which is independent of $J$ and less than that of the fully parametric model (\ref{Eq2.1.1}) with (\ref{Eq2.1.2}) when $J$ is large. Here, $p_{\rm{A}}(p_{\rm{A}}+1)/2$ is the number of parameters included in ${\bm{\Sigma}}_0$.
	
	The simplest model of (\ref{Eq2.1.5}) is the location-shifting MEM with $p_{\rm{A}}=1$ and ${\bm{X}}_{{\rm{A}}ij}\equiv1$, denoted ${\bm{\theta}}_{\rm{A}}^0$ and ${\bm{u}}_j$ by the scalars $\theta_{\rm{A}}^0$ and $u_j$,
	\begin{equation}
		\gamma(u_j, {\bm{x}}_{\rm{B}})=\exp\left[\theta_{\rm{A}}^0+u_j+\left({\bm{\theta}}_{\rm{B}}^0\right)^\top{\bm{x}}_{\rm{B}}\right],\quad j\in\mathcal{J},\label{Eq2.1.6}
	\end{equation}
	which can be regarded as an EVI regression version of the nested error regression model (see, Battese, Harter, and Fuller 1988). The model (\ref{Eq2.1.6}) indicates that the intercept of $\log\gamma$ accepts the heterogeneity between areas, although each covariate has the common slope of $\log\gamma$ across all areas. The nested error regression model is useful for SAE (see, Diallo and Rao 2018; Sugasawa and Kubokawa 2020). The application in Section \ref{Sec4} demonstrates the analysis using the model (\ref{Eq2.1.6}) and verifies its effectiveness numerically. Alternatively, the case $p=p_{\rm{A}}$ yields the most complicated model of (\ref{Eq2.1.5}), indicating that the slope of $\log \gamma$ with respect to each covariate varies across areas.
	
	\subsection{Approximate maximum likelihood estimator}\label{Sec2.2}
	
	In this section, we construct estimators of the unknown parameters $\{{\bm{\theta}}_{\rm{A}}^0, {\bm{\theta}}_{\rm{B}}^0, {\bm{\Sigma}}_0\}$ included in the model (\ref{Eq2.1.4}) with (\ref{Eq2.1.5}).
	
	Let $F_{\omega_j}(y\mid {\bm{u}}_j, {\bm{x}})\coloneqq P(Y_{ij}\leq y\mid {\bm{U}}_j={\bm{u}}_j, {\bm{X}}_{ij}={\bm{x}}, Y_{ij}>\omega_j)$ be the conditional distribution function given ${\bm{U}}_j={\bm{u}}_j$, ${\bm{X}}_{ij}={\bm{x}}$ and $Y_{ij}>\omega_j$, where $\omega_j\in\mathbb{R}^+,\ j\in\mathcal{J}$ are thresholds. In this paper, we assume that $\mathcal{L}(y; {\bm{u}}_j, {\bm{x}})$ belongs to the Hall class (see, Hall 1982), which is mentioned in (A1) of Section \ref{Sec3.1}. From this, we have
	\begin{equation}
		F_{\omega_j}(y\mid {\bm{u}}_j, {\bm{x}})\approx1-\left(\frac{y}{\omega_j}\right)^{-1/\gamma({\bm{u}}_j, {\bm{x}})},\quad j\in\mathcal{J}\label{Eq2.2.1}
	\end{equation}
	for sufficiently large $\omega_j$. Using (\ref{Eq2.2.1}) instead of (\ref{Eq2.1.4}), we can remove ${\mathcal{L}}$ for the estimation of $\gamma$. 
	Similarly, from (\ref{Eq2.2.1}) and the assumption (A1), the density of $Y_{ij}$ given ${\bm{U}}_j={\bm{u}}_j$, ${\bm{X}}_{ij}={\bm{x}}$ and $Y_{ij}>\omega_j$ is obtained as follows:
	\begin{equation}
		f_{w_j}(y\mid {\bm{u}}_j, {\bm{x}})\approx\omega_j^{-1}\gamma({\bm{u}}_j, {\bm{x}})^{-1}\left(\frac{y}{\omega_j}\right)^{-1/\gamma({\bm{u}}_j, {\bm{x}})-1},\quad j\in\mathcal{J}.\label{Eq2.2.2}
	\end{equation}
	Wang and Tsai (2009) considered the similar approximated distribution (\ref{Eq2.2.1}) and density (\ref{Eq2.2.2}) in linear extreme value index regression. Estimation using data exceeding thresholds is the so-called peak-over-threshold method (see, Hill 1975; Wang and Tsai 2009).
	
	We assume that ${\bm{U}}_j$ and ${\bm{X}}_{ij}$ are independent for $i\in\mathcal{N}_j$ and $j\in\mathcal{J}$. Furthermore, we assume that $Y_{ij}$ given ${\bm{U}}_j$ and ${\bm{X}}_{ij}$ is conditionally independent for $i\in\mathcal{N}_j$ and $j\in\mathcal{J}$ and has the distribution function (\ref{Eq2.1.4}) (see, Jiang, Wand, and Bhaskaran 2022). Using (\ref{Eq2.2.2}), we then define the likelihood of $({\bm{\theta}}_{\rm{A}}^0, {\bm{\theta}}_{\rm{B}}^0, {\bm{\Sigma}}_0)$ as
	\begin{equation*}
		L({\bm{\theta}}_{\rm{A}}, {\bm{\theta}}_{\rm{B}}, {\bm{\Sigma}})\coloneqq\prod_{j=1}^JE_{{\bm{U}}_j}\left[\prod_{i\in\mathcal{N}_j: Y_{ij}>\omega_j}f_{w_j}(Y_{ij}\mid {\bm{U}}_j, {\bm{X}}_{ij})\right],
	\end{equation*}
	where $E_{{\bm{U}}_j}$ denotes the expectation over the random effects distribution, $({\bm{\theta}}_{\rm{A}}^\top, {\bm{\theta}}_{\rm{B}}^\top)^\top\in\mathbb{R}^{p_{\rm{A}}}\times\mathbb{R}^{p_{\rm{B}}}$ is any vector corresponding to $(({\bm{\theta}}_{\rm{A}}^0)^\top, ({\bm{\theta}}_{\rm{B}}^0)^\top)^\top$, and ${\bm{\Sigma}}\in\mathbb{R}^{p_{\rm{A}}\times p_{\rm{A}}}$ is any positive definite matrix corresponding to ${\bm{\Sigma}}_0$. The above $L$ is derived from the standard definition of the likelihood for the MEM, because ${\bm{U}}_{j},\ j\in\mathcal{J}$ are unobserved random variables, unlike the data (\ref{Eq2.0.1}) (see, Section 2 of Wu 2009). From (\ref{Eq2.1.3}), (\ref{Eq2.1.5}), (\ref{Eq2.2.2}) and (A1), the log-likelihood of $({\bm{\theta}}_{\rm{A}}^0, {\bm{\theta}}_{\rm{B}}^0, {\bm{\Sigma}}_0)$ can be expressed as
	\begin{align}
		\ell({\bm{\theta}}_{\rm{A}}, {\bm{\theta}}_{\rm{B}}, {\bm{\Sigma}})&\coloneqq\log L({\bm{\theta}}_{\rm{A}}, {\bm{\theta}}_{\rm{B}}, {\bm{\Sigma}})\nonumber\\
		\begin{split}
			&\approx\sum_{j=1}^J\log\int_{\mathbb{R}^{p_{\rm{A}}}}\phi({\bm{u}}; {\bm{0}}, {\bm{\Sigma}})\exp\left(\sum_{i=1}^{n_j}\biggl\{-\left({\bm{\theta}}_{\rm{A}}+{\bm{u}}\right)^\top{\bm{X}}_{{\rm{A}}ij}-{\bm{\theta}}_{\rm{B}}^\top{\bm{X}}_{{\rm{B}}ij}\biggr.\right.\\
			&\quad\Biggl.\left.-\exp\left[-\left({\bm{\theta}}_{\rm{A}}+{\bm{u}}\right)^\top{\bm{X}}_{{\rm{A}}ij}-{\bm{\theta}}_{\rm{B}}^\top{\bm{X}}_{{\rm{B}}ij}\right]\log\frac{Y_{ij}}{\omega_j}\right\}I(Y_{ij}>\omega_j)\Biggr)d{\bm{u}}+C,\label{Eq2.2.3}
		\end{split}
	\end{align}
	where $I(\cdot)$ is an indicator function that returns 1 if $Y_{ij}>\omega_j$ and 0 otherwise, $\phi(\cdot; {\bm{0}}, {\bm{\Sigma}})$ is a density function of $N({\bm{0}}, {\bm{\Sigma}})$, and $C$ is a suitable constant independent of $({\bm{\theta}}_{\rm{A}}, {\bm{\theta}}_{\rm{B}}, {\bm{\Sigma}})$. Again, because ${\bm{U}}_j,\ j\in\mathcal{J}$ are not observed as data, the log-likelihood (\ref{Eq2.2.3}) includes the integral over the domain $\mathbb{R}^{p_{\rm{A}}}$ of the random effects. We denote the approximate maximum likelihood estimator of $({\bm{\theta}}_{\rm{A}}^0, {\bm{\theta}}_{\rm{B}}^0, {\bm{\Sigma}}_0)$ by $(\hat{\bm{\theta}}_{\rm{A}}, \hat{\bm{\theta}}_{\rm{B}}, \hat{\bm{\Sigma}})$, which is the maximizer of the right-hand side of (\ref{Eq2.2.3}).
	
	\subsection{Prediction of random effects}\label{Sec2.3}
	
	In the proposed model (\ref{Eq2.1.4}) using (\ref{Eq2.1.5}), we are not only interested in estimating the parameters $\{{\bm{\theta}}_{\rm{A}}^0, {\bm{\theta}}_{\rm{B}}^0, {\bm{\Sigma}}_0\}$, but also in predicting the random effects ${\bm{U}}_j,\ j\in\mathcal{J}$. Here, we propose the conditional mode method to predict these random effects (see, Santner and Duffy 1989; Section 11 of Wu 2009). Now, the conditional density function of $({\bm{U}}_1, {\bm{U}}_2, \ldots, {\bm{U}}_J)$ given the data (\ref{Eq2.0.1}) is proportional to
	\begin{equation*}
		\prod_{j=1}^J\left[\phi({\bm{u}}_j; {\bm{0}}, {\bm{\Sigma}}_0)\prod_{i\in\mathcal{N}_j: Y_{ij}>\omega_j}f_{\omega_j}(Y_{ij}\mid {\bm{u}}_j, {\bm{X}}_{ij})\right],
	\end{equation*}
	as a function of $({\bm{u}}_1, {\bm{u}}_2, \ldots, {\bm{u}}_J)$. Then, the predictor of ${\bm{U}}_j$ is defined as the mode of this conditional distribution, i.e.,
	\begin{equation}
		\tilde{\bm{u}}_j\coloneqq\operatorname*{argmax}_{{\bm{u}}_j\in\mathbb{R}^{p_{\rm{A}}}}\ \phi({\bm{u}}_j; {\bm{0}}, {\bm{\Sigma}}_0)\prod_{i\in\mathcal{N}_j: Y_{ij}>\omega_j}f_{\omega_j}(Y_{ij}\mid {\bm{u}}_j, {\bm{X}}_{ij}),\quad j\in\mathcal{J},\label{Eq2.3.1}
	\end{equation}
	where $f_{\omega_j}$ and $({\bm{\theta}}_{\rm{A}}^0, {\bm{\theta}}_{\rm{B}}^0, {\bm{\Sigma}}_0)$ included in $f_{\omega_j}$ are replaced by (\ref{Eq2.2.2}) and the estimator $(\hat{\bm{\theta}}_{\rm{A}}, \hat{\bm{\theta}}_{\rm{B}}, \hat{\bm{\Sigma}})$, respectively.
	
	\subsection{Threshold selection}\label{Sec2.4}
	
	The thresholds $\omega_j,\ j\in\mathcal{J}$ in (\ref{Eq2.2.3}) are tuning parameters that balance between the quality of the approximation (\ref{Eq2.2.2}) and the amount of data exceeding the thresholds. By setting higher thresholds, we can generally improve the estimation bias, but the estimator becomes more unstable. Conversely, if we lower the thresholds, the estimator will behave more stably, but may be more biased. Therefore, these thresholds $\omega_j,\ j\in\mathcal{J}$ should be chosen appropriately, considering this trade-off relationship. Here, for each area, we apply the discrepancy measure (see, Wang and Tsai 2009), which considers the goodness of fit of the model, to select the area-wise optimal threshold. In the simulation studies in Section 1.2 of our supplementary material, we verify the performance of the proposed estimator using this threshold selection method.
	
	\section{Asymptotic properties}\label{Sec3}
	
	We investigate the asymptotic properties of the proposed estimator $(\hat{\bm{\theta}}_{\rm{A}}, \hat{\bm{\theta}}_{\rm{B}}, \hat{\bm{\Sigma}})$. In general, the following three types of asymptotic scenarios may be considered:
	
	\renewcommand{\theenumi}{\roman{enumi}}
	\renewcommand{\labelenumi}{(\theenumi)}
	\begin{enumerate}
		\item $J$ remains finite while $n_j,\ j\in\mathcal{J}$ tend to infinity.
		\item $n_j,\ j\in\mathcal{J}$ remain finite while $J$ tends to infinity.
		\item $J$ and $n_j,\ j\in\mathcal{J}$ tend to infinity.
	\end{enumerate}
	In applications using the peak-over-threshold method, the sample size of threshold exceedances is often small for some areas. Therefore, we want to use as many related area sources as possible. Such a scenario can be expressed mathematically as $J\to\infty$. Thus, (i) does not match the background of using the MEM for extreme value analysis. Meanwhile, if the thresholds $\omega_j,\ j\in\mathcal{J}$ as well as the sample sizes exceeding the thresholds are fixed, the consistency of the proposed estimator would not be shown because the bias occurring from the approximation (\ref{Eq2.2.2}) cannot be improved. Ignoring such a bias is outside the concept of EVT (see, Theorems 2 and 4 of Wang and Tsai 2009). This implies that (ii) is also not realistic in our study. To evaluate the impact of the choice of thresholds and bias of the proposed estimator, we must consider the case where $\omega_j\to\infty,\ j\in\mathcal{J}$ and the sample sizes exceeding the thresholds also tend to infinity, which can be taken under $n_j\to\infty,\ j\in\mathcal{J}$. Consequently, (iii) is most important for establishing EVT for the MEM, and we assume this case in the following Sections \ref{Sec3.1} and \ref{Sec3.2}.
	
	Nie (2007) and Jiang, Wand, and Bhaskaran (2022) showed the asymptotic normality of the maximum likelihood estimator of the generalized mixed effects model under (iii). Thus, we can say that the following Theorem \ref{Thm1} extends their results from the generalized mixed effects model to the MEM for EVT.
	
	\subsection{Conditions}\label{Sec3.1}
	
	Let $n_{j0}\coloneqq\sum_{i=1}^{n_j}I(Y_{ij}>\omega_j)$, which is the sample size exceeding the threshold $\omega_j$ for the $j$th area. Additionally, we define $n_0\coloneqq J^{-1}\sum_{j=1}^Jn_{j0}$ as the average of the effective sample sizes of all areas. Note that $n_{j0},\ j\in\mathcal{J}$ and $n_0$ are random variables, not constants. In the case (iii) defined above, we assume that for each $j\in\mathcal{J}$, the threshold $\omega_j$ diverges to infinity in tandem with the sequence of $J$ and the $j$th within-area sample size $n_j$. Accordingly, we denote $\omega_j$ by $\omega_{(J, n_j)}$. The asymptotic properties of the proposed estimator $(\hat{\bm{\theta}}_{\rm{A}}, \hat{\bm{\theta}}_{\rm{B}}, \hat{\bm{\Sigma}})$ rely on the following assumptions (A1)-(A6):
	
	\renewcommand{\theenumi}{A\arabic{enumi}}
	\renewcommand{\labelenumi}{(\theenumi)}
	\begin{enumerate}
		\item $\mathcal{L}(y; {\bm{u}}, {\bm{x}})$ in (\ref{Eq2.1.4}) belongs to the Hall class (see, Hall 1982), that is, 
		\begin{equation}
			\mathcal{L}(y; {\bm{u}}, {\bm{x}})=c_0({\bm{u}}, {\bm{x}})+c_1({\bm{u}}, {\bm{x}})y^{-\beta({\bm{u}}, {\bm{x}})}+\lambda(y; {\bm{u}}, {\bm{x}}),\label{Eq3.1.1}
		\end{equation}
		where $c_0({\bm{u}}, {\bm{x}})>0$, $c_1({\bm{u}}, {\bm{x}})$, $\beta({\bm{u}}, {\bm{x}})>0$ and $\lambda(y; {\bm{u}}, {\bm{x}})$ are continuous and bounded. Furthermore, $\lambda(y; {\bm{u}}, {\bm{x}})$ satisfies
		\begin{equation*}
			\sup_{{\bm{u}}\in\mathbb{R}^{p_{\rm{A}}},\ {\bm{x}}\in\mathbb{R}^p}\left[y^{\beta({\bm{u}}, {\bm{x}})}\lambda(y; {\bm{u}}, {\bm{x}})\right]\to0\quad{\rm{as}}\quad y\to\infty.
		\end{equation*}
		\item There exists a bounded and continuous function $\delta: \mathbb{R}^{p_{\rm{A}}}\times\mathbb{R}^p\to\mathbb{R}^+$ such that
		\begin{equation*}
			\sup_{{\bm{u}}\in\mathbb{R}^{p_{\rm{A}}},\ {\bm{x}}\in\mathbb{R}^p}\left\lvert\frac{P(Y_{ij}>y\mid {\bm{U}}_j={\bm{u}}, {\bm{X}}_{ij}={\bm{x}})}{P(Y_{ij}>y\mid {\bm{U}}_j={\bm{u}})}-\delta({\bm{u}}, {\bm{x}})\right\lvert\to0\quad{\rm{as}}\quad y\to\infty.
		\end{equation*}
		\item As $n_j\to\infty,\ j\in\mathcal{J}$ and $J\to\infty$,
		\begin{equation*}
			\inf_{j\in\mathcal{J},\ {\bm{u}}\in\mathbb{R}^{p_{\rm{A}}}}n_jP(Y_{ij}>\omega_{(J, n_j)}\mid {\bm{U}}_j={\bm{u}})\to\infty.
		\end{equation*}
		\item There exist some bounded and continuous functions $d_j: \mathbb{R}^{p_{\rm{A}}}\to\mathbb{R}^+,\ j\in\mathcal{J}$ such that under given ${\bm{U}}_j={\bm{u}}$, $n_{j0}/n_0\to^Pd_j({\bm{u}})$ uniformly for all $j\in\mathcal{J}$ and ${\bm{u}}\in\mathbb{R}^{p_{\rm{A}}}$ as $n_j\to\infty,\ j\in\mathcal{J}$ and $J\to\infty$, where the symbol ``$\to^P$'' represents convergence in probability.
		\item $n_0/J\to^P0$ as $n_j\to\infty,\ j\in\mathcal{J}$ and $J\to\infty$.
		\item There exist some bounded and continuous functions ${\bm{b}}_{{\rm{K}}j}: \mathbb{R}^{p_{\rm{A}}}\to\mathbb{R},\ j\in\mathcal{J},\ {\rm{K}}\in\{{\rm{A}}, {\rm{B}}\}$ such that
		\begin{equation*}
			\sup_{j\in\mathcal{J},\ {\bm{u}}\in\mathbb{R}^{p_{\rm{A}}}}\left\lVert\frac{J^{1/2}n_0^{1/2}{E_{{\bm{X}}_{ij}}}\left[{\bm{X}}_{{\rm{K}}ij}\zeta_j({\bm{u}}, {\bm{X}}_{ij})\right]}{P(Y_{ij}>\omega_{(J, n_j)}\mid {\bm{U}}_j={\bm{u}})^{1/2}}-{\bm{b}}_{{\rm{K}}j}({\bm{u}})\right\rVert\to0\\
		\end{equation*}
		as $n_j\to\infty,\ j\in\mathcal{J}$ and $J\to\infty$, where
		\begin{equation*}
			\zeta_j({\bm{u}}, {\bm{x}})\coloneqq\frac{c_1({\bm{u}}, {\bm{x}})\gamma({\bm{u}}, {\bm{x}})\beta({\bm{u}}, {\bm{x}})}{1+\gamma({\bm{u}}, {\bm{x}})\beta({\bm{u}}, {\bm{x}})}\omega_{(J, n_j)}^{-1/\gamma({\bm{u}}, {\bm{x}})-\beta({\bm{u}}, {\bm{x}})}
		\end{equation*}
		and $\lVert\cdot\rVert$ refers to the Euclidean norm.
	\end{enumerate}
	(A1) and (A2) regularize the tail behavior of the conditional response distribution (\ref{Eq2.1.4}) (see, Wang and Tsai 2009; Ma, Jiang, and Huang 2019). (A3)-(A6) impose the constraints on the divergence rates of the thresholds $\omega_{(J, n_j)},\ j\in\mathcal{J}$. (A3) implies that for each $j\in\mathcal{J}$, the effective sample size $n_{j0}$ asymptotically diverges to infinity. Under (A4), $n_{j0},\ j\in\mathcal{J}$ are not critically different. Furthermore, (A5) means that the number of areas $J$ is relatively larger than the effective sample sizes $n_{j0},\ j\in\mathcal{J}$. (A5) mathematically links the divergence rates of $\omega_{(J, n_j)},\ j\in\mathcal{J}$ and $J$. (A6) is related to the asymptotic bias of the proposed estimator. If (A6) fails, the consistency of the proposed estimator may not be guaranteed.
	
	\subsection{Asymptotic normality}\label{Sec3.2}
	
	Let ${\bm{M}}$ be a matrix of zeros and ones such that ${\bm{M}}{\rm{vech}}({\bm{A}})={\rm{vec}}({\bm{A}})$ for all symmetric matrices ${\bm{A}}\in\mathbb{R}^{p_{\rm{A}}\times p_{\rm{A}}}$, where ${\rm{vec}}(\cdot)$ is a vector operator, and ${\rm{vech}}(\cdot)$ is a vector half operator that stacks the lower triangular half of a given $d\times d$ square matrix into the single vector of length $d(d+1)/2$ (see, Magnus and Neudecker 1988). The Moore-Penrose inverse of ${\bm{M}}$ is ${\bm{M}}_*\coloneqq({\bm{M}}^\top{\bm{M}})^{-1}{\bm{M}}^\top$.
	
	For the maximum likelihood estimator $(\hat{\bm{\theta}}_{\rm{A}}, \hat{\bm{\theta}}_{\rm{B}}, \hat{\bm{\Sigma}})$, we obtain the following Theorem \ref{Thm1}.
	
	\begin{theorem}\label{Thm1}
		Suppose that (A1)-(A6) hold.
		Then, as $n_j\to\infty,\ j\in\mathcal{J}$ and $J\to\infty$,
		\begin{equation*} 
			\begin{bmatrix}
				J^{1/2}\left(\hat{\bm{\theta}}_{\rm{A}}-{\bm{\theta}}_{\rm{A}}^0\right)\\
				J^{1/2}n_0^{1/2}\left(\hat{\bm{\theta}}_{\rm{B}}-{\bm{\theta}}_{\rm{B}}^0\right)\\
				J^{1/2}{\rm{vech}}\left(\hat{\bm{\Sigma}}-{\bm{\Sigma}}_0\right)
			\end{bmatrix}+
			\begin{bmatrix}
				n_0^{-1/2}{\bm{b}}_{\rm{A}}\\
				{\bm{b}}_{\rm{B}}\\
				n_0^{-1/2}{\bm{b}}_{\rm{C}}\\
			\end{bmatrix}
			\xrightarrow{D}N\left({\bm{0}}, 
			\begin{bmatrix}
				{\bm{\Delta}}_{\rm{A}} & {\bm{O}} & {\bm{O}}\\
				{\bm{O}} & {\bm{\Delta}}_{\rm{B}} & {\bm{O}}\\
				{\bm{O}} & {\bm{O}} & {\bm{\Delta}}_{\rm{C}}
			\end{bmatrix}\right),
		\end{equation*}
		where the symbol ``$\to^D$'' denotes convergence in distribution, ${\bm{O}}$s are zero matrices of appropriate size, and ${\bm{b}}_{\rm{K}}$ and ${\bm{\Delta}}_{\rm{K}},\ {\rm{K}}\in\{{\rm{A}}, {\rm{B}}, {\rm{C}}\}$ are defined as follows:
		\begin{align*}
			{\bm{b}}_{\rm{A}}&\coloneqq\lim_{J\to\infty}J^{-1}\sum_{j=1}^JE\left[d_j({\bm{U}}_j)^{-1/2}{\bm{\Phi}}_{\rm{AA}}({\bm{U}}_j)^{-1}{\bm{b}}_{{\rm{A}}j}({\bm{U}}_j)\right],\\
			{\bm{b}}_{\rm{B}}&\coloneqq\lim_{J\to\infty}J^{-1}\sum_{j=1}^J{\bm{\Delta}}_{\rm{B}}E\left[d_j({\bm{U}}_j)^{1/2}\left[{\bm{b}}_{{\rm{B}}j}({\bm{U}}_j)-{\bm{\Phi}}_{\rm{AB}}({\bm{U}}_j)^\top{\bm{\Phi}}_{\rm{AA}}({\bm{U}}_j)^{-1}{\bm{b}}_{{\rm{A}}j}({\bm{U}}_j)\right]\right],\\
			\begin{split}
				{\bm{b}}_{\rm{C}}&\coloneqq\lim_{J\to\infty}J^{-1}\sum_{j=1}^J{\bm{\Delta}}_{\rm{C}}{\bm{M}}_*\left({\bm{\Sigma}}_0\otimes{\bm{\Sigma}}_0\right)^{-1}\\
				&\quad\times{\rm{vec}}\left(E\left[d_j({\bm{U}}_j)^{-1/2}\left[{\bm{U}}_j{\bm{b}}_{{\rm{A}}j}({\bm{U}}_j)^\top{\bm{\Phi}}_{\rm{AA}}({\bm{U}}_j)^{-1}+{\bm{\Phi}}_{\rm{AA}}({\bm{U}}_j)^{-1}{\bm{b}}_{{\rm{A}}j}({\bm{U}}_j){\bm{U}}_j^\top\right]\right]\right),
			\end{split}\\
			{\bm{\Delta}}_{\rm{A}}&\coloneqq{\bm{\Sigma}}_0,\\
			{\bm{\Delta}}_{\rm{B}}&\coloneqq E\left[{\bm{\Phi}}_{\rm{BB}}({\bm{U}}_j)-{\bm{\Phi}}_{\rm{AB}}({\bm{U}}_j)^\top{\bm{\Phi}}_{\rm{AA}}({\bm{U}}_j)^{-1}{\bm{\Phi}}_{\rm{AB}}({\bm{U}}_j)\right]^{-1}\quad{\text{and}}\\
			{\bm{\Delta}}_{\rm{C}}&\coloneqq2\left[{\bm{M}}_*\left({\bm{\Sigma}}_0\otimes{\bm{\Sigma}}_0\right)^{-1}{\bm{M}}_*^\top\right]^{-1},
		\end{align*}
		where ${\bm{\Phi}}_{{\rm{K}}_1{\rm{K}}_2}({\bm{U}}_j)\coloneqq E_{{\bm{X}}_{ij}}[\delta({\bm{U}}_j, {\bm{X}}_{ij}){\bm{X}}_{{\rm{K}}_1ij}{\bm{X}}_{{\rm{K}}_2ij}^\top]$ for ${\rm{K}}_1, {\rm{K}}_2\in\{{\rm{A}}, {\rm{B}}\}$, and $\otimes$ is the Kronecker product.
	\end{theorem}

\bigskip

	\begin{remark}\label{Rem1}
		From Theorem \ref{Thm1}, $\hat{\bm{\theta}}_{\rm{A}}$ and $\hat{\bm{\Sigma}}$ are $\sqrt{J}$-consistent, and $\hat{\bm{\theta}}_{\rm{B}}$ is $\sqrt{Jn_0}$-consistent. Furthermore, $\hat{\bm{\theta}}_{\rm{A}}$, $\hat{\bm{\theta}}_{\rm{B}}$ and $\hat{\bm{\Sigma}}$ are asymptotically independent. If $J$ and $n_j,\ j\in\mathcal{J}$ are sufficiently large, the covariance matrix of the proposed estimator is obtained as
		\begin{equation*}
			{\rm{cov}}\left[\hat{\bm{\theta}}_{\rm{A}}\right]\approx J^{-1}{\bm{\Delta}}_{\rm{A}},\;{\rm{cov}}\left[\hat{\bm{\theta}}_{\rm{B}}\right]\approx (Jn_0)^{-1}{\bm{\Delta}}_{\rm{B}}\quad{\rm{and}}\quad{\rm{cov}}\left[{\rm{vech}}\left(\hat{\bm{\Sigma}}\right)\right]\approx J^{-1}{\bm{\Delta}}_{\rm{C}}.
		\end{equation*}
		Theorem \ref{Thm1} also reveals the asymptotic bias of the proposed estimator caused by the approximation (\ref{Eq2.2.2}). If $J$ and $n_j,\ j\in\mathcal{J}$ are sufficiently large, it can be approximated as
		\begin{align*}
			&E\left[\hat{\bm{\theta}}_{\rm{A}}\right]-{\bm{\theta}}_{\rm{A}}^0\approx\left(Jn_0\right)^{-1/2}{\bm{b}}_{\rm{A}},\\
			&E\left[\hat{\bm{\theta}}_{\rm{B}}\right]-{\bm{\theta}}_{\rm{B}}^0\approx\left(Jn_0\right)^{-1/2}{\bm{b}}_{\rm{B}}\quad{\rm{and}}\\
			&E\left[{\rm{vech}}\left(\hat{\bm{\Sigma}}\right)\right]-{\rm{vech}}\left({\bm{\Sigma}}_0\right)\approx\left(Jn_0\right)^{-1/2}{\bm{b}}_{\rm{C}}.
		\end{align*}
		As shown in (A6), ${\bm{b}}_{{\rm{K}}j}$ depends on the EVI function $\gamma({\bm{u}}, {\bm{x}})$, and the proposed estimator is more biased for larger $\gamma({\bm{u}}, {\bm{x}})$, that is, the heavier the right tail of the response distribution (\ref{Eq2.1.4}). Furthermore, ${\bm{b}}_{{\rm{K}}j}$ is also affected by $\beta({\bm{u}}, {\bm{x}})$ defined in (\ref{Eq3.1.1}), and the proposed estimator is more biased for smaller $\beta({\bm{u}}, {\bm{x}})$. Meanwhile, $c_0({\bm{u}}, {\bm{x}})$ in (\ref{Eq3.1.1}), which is the scaling constant to ensure that the upper bound of (\ref{Eq2.1.4}) is equal to one, is not related to the asymptotic bias of the proposed estimator.
	\end{remark}

\bigskip

	\begin{remark}\label{Rem2}
		From Theorem \ref{Thm1}, we can confirm the good compatibility between the MEM and EVT as follows. In extreme value analysis, we want to set the threshold as high as possible to ensure a good fit with the Pareto distribution, as shown in (\ref{Eq2.2.2}). However, the estimator may have a large variance because the amount of available data becomes small. Meanwhile, the variance of the proposed estimator for (\ref{Eq2.1.4}) with (\ref{Eq2.1.5}) depends strongly on the number of areas $J$ and improves as $J$ increases, as described in Remark \ref{Rem1}. Note that the magnitude of $J$ is unaffected by the choice of thresholds $\omega_j,\ j\in\mathcal{J}$, unlike $n_0$. Therefore, even if the threshold is high for some areas, the proposed estimator is expected to remain stable as long as $J$ is sufficiently large. Note that estimating the bias of the proposed estimator is a difficult problem because $\beta({\bm{u}}, {\bm{x}})$ and $c_1({\bm{u}}, {\bm{x}})$ in (\ref{Eq3.1.1}) must be estimated. However, if $J$ is sufficiently large, by setting reasonably high thresholds, we may avoid this bias estimation problem while ensuring the stability of the estimator. Such phenomena are numerically confirmed in the simulation study in Section 1.2 of our supplementary material.
	\end{remark}

\bigskip

	\begin{remark}\label{Rem3}
		Theorem \ref{Thm1} is directly applicable to confidence interval construction and statistical hypothesis testing on the parameters ${\bm{\theta}}_{\rm{A}}^0$, ${\bm{\theta}}_{\rm{B}}^0$ and ${\bm{\Sigma}}_0$. To obtain more efficient estimates, the choice of covariates is crucial. Alternatively, including too many meaningless covariates in the model will adversely affect the parameter estimates, and ``borrowing of strength'' will not be effective. Therefore, we must check the efficiency of the selected explanatory variables. Hypothesis testing is useful for this purpose. The typical statement of such a hypothesis test is whether or not each component of ${\bm{\theta}}_{\rm{A}}^0$ and ${\bm{\theta}}_{\rm{B}}^0$ is significantly different from zero. When we organize this test, we must estimate ${\bm{\Delta}}_{\rm{B}}^{-1}$, which can be naturally estimated by
		\begin{equation}
			\hat{\bm{\Delta}}_{\rm{B}}^{-1}\coloneqq J^{-1}\sum_{j=1}^J\left(\hat{\bm{\Phi}}_{{\rm{BB}}j}-\hat{\bm{\Phi}}_{{\rm{AB}}j}^\top\hat{\bm{\Phi}}_{{\rm{AA}}j}^{-1}\hat{\bm{\Phi}}_{{\rm{AB}}j}\right),\label{Eq3.2.1}
		\end{equation}
		where $\hat{\bm{\Phi}}_{{\rm{K}}_1{\rm{K}}_2j}\coloneqq n_{j0}^{-1}\sum_{i=1}^{n_j}{\bm{X}}_{{\rm{K}}_1ij}{\bm{X}}_{{\rm{K}}_2ij}^\top I(Y_{ij}>\omega_j),\ {\rm{K}}_1, {\rm{K}}_2\in\{{\rm{A}}, {\rm{B}}\}$. In Section \ref{Sec4.3}, the hypothesis test on ${\bm{\theta}}_{\rm{B}}^0$ is demonstrated for a real dataset (see, Section 1.1 of our supplementary material).
	\end{remark}
	
	\bigskip
	
	As described in Section \ref{Sec2.1}, an important example of (\ref{Eq2.1.5}) is the type of nested error regression model (\ref{Eq2.1.6}). For the model (\ref{Eq2.1.6}), Theorem \ref{Thm1} can be simplified to the following Corollary \ref{Cor1}. Let define $\sigma_0^2\coloneqq{\rm{var}}[U_j]$ for the random effects $U_j,\ j\in\mathcal{J}$ and denote its proposed estimator as $\hat{\sigma}^2$.
	
	\begin{corollary}\label{Cor1}
		Suppose that (A1)-(A6) hold.
		Then, as $n_j\to\infty,\ j\in\mathcal{J}$ and $J\to\infty$,
		\begin{equation*} 
			\begin{bmatrix}
				J^{1/2}\left(\hat{\theta}_{\rm{A}}-\theta_{\rm{A}}^0\right)\\
				J^{1/2}n_0^{1/2}\left(\hat{\bm{\theta}}_{\rm{B}}-{\bm{\theta}}_{\rm{B}}^0\right)\\
				J^{1/2}\left(\hat{\sigma}^2-\sigma_0^2\right)
			\end{bmatrix}+
			\begin{bmatrix}
				n_0^{-1/2}v_{\rm{A}}\\
				{\bm{v}}_{\rm{B}}\\
				n_0^{-1/2}v_{\rm{C}}\\
			\end{bmatrix}
			\xrightarrow{D}N\left({\bm{0}}, 
			\begin{bmatrix}
				\sigma_0^2 & {\bm{O}} & {\bm{O}}\\
				{\bm{O}} & {\bm{\Omega}}_{\rm{B}} & {\bm{O}}\\
				{\bm{O}} & {\bm{O}} & 2\left(\sigma_0^2\right)^2
			\end{bmatrix}\right),
		\end{equation*}
		where $v_{\rm{A}}$, ${\bm{v}}_{\rm{B}}$, $v_{\rm{C}}$ and ${\bm{\Omega}}_{\rm{B}}$ are defined as follows:
		\begin{align*}
			&v_{\rm{A}}\coloneqq\lim_{J\to\infty}J^{-1}\sum_{j=1}^JE\left[d_j(U_j)^{-1/2}v_{{\rm{A}}j}(U_j)\right],\\
			&{\bm{v}}_{\rm{B}}\coloneqq\lim_{J\to\infty}J^{-1}\sum_{j=1}^J{\bm{\Omega}}_{\rm{B}}E\left[d_j(U_j)^{1/2}\left[{\bm{b}}_{{\rm{B}}j}({\bm{U}}_j)-v_{{\rm{A}}j}(U_j){\bm{\Psi}}_{\rm{B}}(U_j)\right]\right],\\
			&v_{\rm{C}}\coloneqq\lim_{J\to\infty}J^{-1}\sum_{j=1}^J4{\rm{vec}}\left(E\left[d_j(U_j)^{-1/2}v_{{\rm{A}}j}(U_j)U_j\right]\right)\quad{\text{and}}\\
			&{\bm{\Omega}}_{\rm{B}}\coloneqq E\left[{\bm{\Phi}}_{\rm{BB}}(U_j)-{\bm{\Psi}}_{\rm{B}}(U_j){\bm{\Psi}}_{\rm{B}}(U_j)^\top\right]^{-1},
		\end{align*}
		where $v_{{\rm{A}}j}(U_j)$ is ${\bm{b}}_{{\rm{A}}j}(U_j)$ with $p_{\rm{A}}=1$ and ${\bm{X}}_{{\rm{A}}ij}\equiv1$, ${\bm{\Psi}}_{\rm{B}}(U_j)\coloneqq E_{{\bm{X}}_{{\rm{B}}ij}}[\delta(U_j, {\bm{X}}_{{\rm{B}}ij}){\bm{X}}_{{\rm{B}}ij}]$, and ${\bm{\Phi}}_{\rm{BB}}(U_j)$ is defined in Theorem \ref{Thm1}.
	\end{corollary}
	
	\section{Application}\label{Sec4}
	\subsection{Background}\label{Sec4.1}
	
	In this section, we analyze the returns of a large number of cryptocurrencies. Over the past decade, the cryptocurrency market has grown rapidly, led by Bitcoin. As of January 2024, the number of active cryptocurrency stocks is estimated to be over 9,000 (see, \url{https://www.statista.com/statistics/863917/number-crypto-coins-tokens/}). However, despite the large number of stocks, most studies are limited to a few major cryptocurrencies such as Bitcoin. Gkillas and Katsiampa (2018) studied the extreme value analysis of the returns of five cryptocurrencies, namely Bitcoin, Ethereum, Ripple, Litecoin and Bitcoin Cash. We extend their work to a larger number of stocks including these five cryptocurrencies. Our goal is to demonstrate the advantage of simultaneously analyzing many cryptocurrency stocks using the MEM. In Section \ref{Sec4.3}, we report the interpretation of our model. In Section \ref{Sec4.4}, we compare the performance of our method with that of stock-by-stock analysis using the method of Wang and Tsai (2009).
	
	\subsection{Dataset}\label{Sec4.2}
	
	The cryptocurrency dataset is available on CoinMarketCap (see, \url{https://coinmarketcap.com/}). We use this dataset from the last ten years, from January 1, 2014 to December 31, 2023. Then, we cover the 413 stocks that are in the top 500 on CoinMarketCap as of February 2, 2024 and have returns of 364 days or more. From the daily closing prices for each stock, the returns can be calculated as $\log P_{tj} - \log P_{(t-1)j}$, where $P_{tj}$ refers to the closing price on the $t$th day in the $j$th cryptocurrency. Let the non-missing returns of the $j$th cryptocurrency be denoted as $\{Y_{ij},\ i=1, 2, \ldots, n_j\}$ for $j=1, 2, \ldots, 413$. The sample size $n_j$ of the $j$th cryptocurrency is primarily determined by the launch date. Note that the index $i$ does not represent the same date for all stocks.
	
	We examine the tail behavior of both the negative and positive returns for each stock. In many applications of EVT, the sample kurtosis has been used to check the effectiveness of using the Pareto-type distribution (see, Wang and Tsai 2009; Ma, Jiang, and Huang 2019). In this study, the sample kurtosis of $\{Y_{ij}\}_{i\in\mathcal{N}_j}$ was greater than zero for each $j=1, 2, \ldots, 413$, and the minimum sample kurtosis for the 413 stocks was 1.833, implying that the returns of each cryptocurrency are heavily distributed. Therefore, we use the Pareto-type distribution instead of the GPD to analyze high threshold exceedances for the both negative and positive returns. The following Sections \ref{Sec4.3} and \ref{Sec4.4} describe our method only for positive returns. However, using the same method by replacing $\{Y_{ij}\}_{i\in\mathcal{N}_j, j\in\mathcal{J}}$ with $\{-Y_{ij}\}_{i\in\mathcal{N}_j, j\in\mathcal{J}}$, we can also implement the analysis of extreme negative returns.
	
	For each pair of the 413 stocks, we estimated the tail dependence parameter from returns above the 95th percentile for the both negative and positive returns (see, Reiss and Thomas 2007). The results showed that the percentage of combinations with a tail dependence over 0.7 was only 0.074\% for negative returns and only 0.032\% for positive returns. Therefore, in this analysis, we do not consider the dependence between each pair of stocks.
	
	\subsection{Analysis by our model}\label{Sec4.3}
	
	In many cryptocurrency applications, dummy variables such as year, month, and day of the week have often been utilized as covariates to account for the non-stationarity of returns. In Longin and Pagliardi (2016) and Gkillas and Katsiampa (2018), the returns were adjusted in terms of variance using some dummy variables and were analyzed under the assumption that the EVI of each cryptocurrency remained constant over the period, despite the dramatic growth of the market. However, in their approach, it is unclear which variable influences returns to what extent. Therefore, in our application, we employ EVI regression to examine the effects of dummy variables with year (9 variables), month (11 variables) and day of the week (6 variables) on the tail behavior of the distribution of returns. The second column of Table \ref{Tab1} shows the assignment of 26 dummy variables. We denote the vector of these dummy variables as ${\bm{X}}_{ij}\in\prod_{k=1}^{26}\{0, 1\}\subset\mathbb{R}^{26}$ for $i=1, 2, \ldots, n_j$ and $j=1, 2, \ldots, 413$. For our model in Section \ref{Sec2.1}, we set ${\bm{X}}_{{\rm{A}}ij}\equiv1$ and ${\bm{X}}_{{\rm{B}}ij}={\bm{X}}_{ij}$ and denote the random effects as $U_1, U_2, \ldots, U_{413}\stackrel{\rm{i.i.d.}}{\sim}N(0, \sigma^2)$, where $\sigma^2>0$ is an unknown variance. Then, for the underlying Pareto-type distribution (\ref{Eq2.1.4}), the EVI (\ref{Eq2.1.5}) conditional on $U_j=u_j$ and ${\bm{X}}_{ij}={\bm{x}}$ is modeled as
	\begin{equation}
		\gamma(u_j, {\bm{x}})=\exp\left(\theta_{\rm{A}}+u_j+{\bm{\theta}}_{\rm{B}}^\top{\bm{x}}\right),\quad j=1, 2, \ldots, 413, \label{Eq4.3.1}
	\end{equation}
	where $\theta_{\rm{A}}\in\mathbb{R}$ and ${\bm{\theta}}_{\rm{B}}\in\mathbb{R}^{26}$ are unknown coefficients. Below, we analyze the returns of the 413 cryptocurrencies simultaneously using (\ref{Eq4.3.1}). For the implementation, we then use the \texttt{glmer()} function in the package \textsf{lme4} (see, Bates et al. 2015) within the \textsf{R} computing environment (see, R Core Team 2021). A more detailed explanation is given in Section 1 of our supplementary material.
	
	\begin{table}[t]
		\captionsetup{width=0.95\linewidth}
		\caption{The estimation and test results for the cryptocurrency dataset. In the third and forth columns, the values in parentheses show the widths of the $95\%$ confidence intervals. In the fifth and sixth columns, ``1'' indicates that the null hypothesis was rejected.}
		\label{Tab1}
		\centering
		\scalebox{0.8}{\begin{tabular}{|c|c||rc|rc|c|c|c|c|}\hline
				\multicolumn{2}{|c||}{} & \multicolumn{4}{c|}{Estimate} & \multicolumn{2}{c|}{Rejection} & \multicolumn{2}{c|}{p-value} \\\cline{3-10}
				\multicolumn{2}{|c||}{} & \multicolumn{2}{c|}{Negative} & \multicolumn{2}{c|}{Positive} & Negative & Positive & Negative & Positive \\\hline\hline
				\multicolumn{2}{|c||}{$\theta_{\rm{A}}$} & $-1.337$ & $(0.026)$ & $-0.854$ & $(0.025)$ & $-$ & $-$ & $-$ & $-$ \\\hline
				\multirow{28}{*}{${\bm{\theta}}_{\rm{B}}$} & 2014 & $0.690$ & $(0.167)$ & $0.534$ & $(0.164)$ & 1 & 1 & $<10^{-3}$ & $<10^{-3}$ \\
				& 2015 & $0.732$ & $(0.168)$ & $0.489$ & $(0.167)$ & 1 & 1 & $<10^{-3}$ & $<10^{-3}$ \\
				& 2016 & $0.735$ & $(0.133)$ & $0.603$ & $(0.121)$ & 1 & 1 & $<10^{-3}$ & $<10^{-3}$ \\
				& 2017 & $0.677$ & $(0.070)$ & $0.605$ & $(0.060)$ & 1 & 1 & $<10^{-3}$ & $<10^{-3}$ \\
				& 2018 & $0.461$ & $(0.049)$ & $0.264$ & $(0.050)$ & 1 & 1 & $<10^{-3}$ & $<10^{-3}$ \\
				& 2019 & $0.307$ & $(0.049)$ & $0.165$ & $(0.046)$ & 1 & 1 & $<10^{-3}$ & $<10^{-3}$ \\
				& 2020 & $0.483$ & $(0.040)$ & $0.219$ & $(0.035)$ & 1 & 1 & $<10^{-3}$ & $<10^{-3}$ \\
				& 2021 & $0.383$ & $(0.032)$ & $0.282$ & $(0.027)$ & 1 & 1 & $<10^{-3}$ & $<10^{-3}$ \\
				& 2022 & $0.231$ & $(0.029)$ & $0.018$ & $(0.027)$ & 1 & 0 & $<10^{-3}$ & $0.092$ \\
				& 2023 & \multicolumn{1}{c}{$-$} & \multicolumn{1}{c|}{$-$} & \multicolumn{1}{c}{$-$} & \multicolumn{1}{c|}{$-$} & $-$ & $-$ & $-$ & $-$ \\\cline{2-10}
				& Jan & $0.189$ & $(0.043)$ & $0.050$ & $(0.039)$ & 1 & 1 & $<10^{-3}$ & $0.006$ \\
				& Feb & $0.097$ & $(0.043)$ & $0.095$ & $(0.039)$ & 1 & 1 & $<10^{-3}$ & $<10^{-3}$ \\
				& Mar & $0.141$ & $(0.044)$ & $0.075$ & $(0.038)$ & 1 & 1 & $<10^{-3}$ & $<10^{-3}$ \\
				& Apr & $0.016$ & $(0.043)$ & $0.014$ & $(0.042)$ & 0 & 0 & $0.229$ & $0.257$ \\
				& May & $0.352$ & $(0.038)$ & $0.087$ & $(0.039)$ & 1 & 1 & $<10^{-3}$ & $<10^{-3}$ \\
				& Jun & $0.177$ & $(0.039)$ & $-0.076$ & $(0.042)$ & 1 & 1 & $<10^{-3}$ & $<10^{-3}$ \\
				& Jul & $-0.056$ & $(0.047)$ & $-0.063$ & $(0.041)$ & 1 & 1 & $0.010$ & $0.001$ \\
				& Aug & $-0.069$ & $(0.044)$ & $0.029$ & $(0.041)$ & 1 & 0 & $0.001$ & $0.084$ \\
				& Sep & $0.188$ & $(0.042)$ & $-0.041$ & $(0.042)$ & 1 & 0 & $<10^{-3}$ & $0.028$ \\
				& Oct & $-0.071$ & $(0.048)$ & $-0.039$ & $(0.042)$ & 1 & 0 & $0.002$ & $0.033$ \\
				& Nov & $0.199$ & $(0.040)$ & $0.061$ & $(0.037)$ & 1 & 1 & $<10^{-3}$ & $0.001$ \\
				& Dec & \multicolumn{1}{c}{$-$} & \multicolumn{1}{c|}{$-$} & \multicolumn{1}{c}{$-$} & \multicolumn{1}{c|}{$-$} & $-$ & $-$ & $-$ & $-$ \\\cline{2-10}
				& San & $-0.032$ & $(0.038)$ & $-0.024$ & $(0.034)$ & 0 & 0 & $0.053$ & $0.079$ \\
				& Mon & $0.062$ & $(0.034)$ & $0.022$ & $(0.032)$ & 1 & 0 & $<10^{-3}$ & $0.089$ \\
				& Tue & $0.066$ & $(0.036)$ & $-0.013$ & $(0.033)$ & 1 & 0 & $<10^{-3}$ & $0.212$ \\
				& Wed & $0.183$ & $(0.035)$ & $-0.026$ & $(0.031)$ & 1 & 0 & $<10^{-3}$ & $0.05$ \\
				& Thu & $0.065$ & $(0.035)$ & $0.061$ & $(0.032)$ & 1 & 1 & $<10^{-3}$ & $<10^{-3}$ \\
				& Fri & $0.013$ & $(0.035)$ & $-0.006$ & $(0.032)$ & 0 & 0 & $0.234$ & $0.365$ \\
				& Sat & \multicolumn{1}{c}{$-$} & \multicolumn{1}{c|}{$-$} & \multicolumn{1}{c}{$-$} & \multicolumn{1}{c|}{$-$} & $-$ & $-$ & $-$ & $-$ \\\hline
				\multicolumn{2}{|c||}{$\sigma^2$} & $0.074$ & $(0.010)$ & $0.068$ & $(0.009)$ & $-$ & $-$ & $-$ & $-$ \\\hline
		\end{tabular}}
	\end{table}
	
	First, we estimate the parameters $\{\theta_{\rm{A}}, {\bm{\theta}}_{\rm{B}}, \sigma^2\}$ in (\ref{Eq4.3.1}). The third and fourth columns of Table \ref{Tab1} showed the maximum likelihood estimates of these parameters for negative returns and positive returns, respectively. In these columns, the values in parentheses show the widths of the $95\%$ confidence intervals calculated from Corollary \ref{Cor1}. According to the estimates in Table \ref{Tab1}, for any combination of our dummy variables, the EVI (\ref{Eq4.3.1}) was higher for positive returns than for negative returns. This result suggests that positive returns may have had a greater impact on the cryptocurrency market than negative returns. Accordingly, this seems to support the evidence of the overall growth of the market. Meanwhile, the effects of dummy variables for year reduced the EVI over the years, implying that many cryptocurrency stocks were less risky as assets than in the past.
	
	Second, we conduct the Wald hypothesis tests to check whether each of our covariates has a significant effect in the model (\ref{Eq4.3.1}). The hypothesis tests of interest are expressed as follows:
	\begin{equation*}
		{\rm{H}}_{0k}: ({\bm{\theta}}_{\rm{B}})_k=0\quad{\rm{vs.}}\quad{\rm{H}}_{1k}: ({\bm{\theta}}_{\rm{B}})_k\neq0
	\end{equation*}
	for $k=1, 2, \ldots, 26$, where ${\rm{H}}_{0k}$ is the null hypothesis, ${\rm{H}}_{1k}$ is the alternative hypothesis, and $({\bm{\theta}}_{\rm{B}})_k$ is the $k$th component of ${\bm{\theta}}_{\rm{B}}$. From Corollary \ref{Cor1}, we define the test statistic as
	\begin{equation}
		T_k\coloneqq({\bm{\Omega}}_{\rm{B}})^{-1/2}_{k}\left[(Jn_0)^{1/2}(\hat{\bm{\theta}}_{\rm{B}})_k-({\bm{v}}_{\rm{B}})_k\right],\quad k=1, 2, \ldots, 26,\label{Eq4.3.2}
	\end{equation}
	where $({\bm{\Omega}}_{\rm{B}})_k$ is the $(k, k)$ entry of ${\bm{\Omega}}_{\rm{B}}$, and $(\hat{\bm{\theta}}_{\rm{B}})_k$ and $({\bm{v}}_{\rm{B}})_k$ are the $k$th components of $\hat{\bm{\theta}}_{\rm{B}}$ and ${\bm{v}}_{\rm{B}}$, respectively. In (\ref{Eq4.3.2}), we estimate ${\bm{\Omega}}_{\rm{B}}^{-1}$ by (\ref{Eq3.2.1}) and assume that ${\bm{v}}_{\rm{B}}$ is a zero vector. Under the null hypothesis ${\rm{H}}_{0k}$, the distribution of $T_k$ can be approximated by $N(0, 1)$. Therefore, for a given significance level $\alpha$, we reject the null hypothesis ${\rm{H}}_{0k}$ if $|T_k|>z_{1-\alpha/2}$, where $z_{1-\alpha/2}$ is the $100(1-\alpha/2)$-th percentile point of $N(0, 1)$. The fifth and sixth columns of Table \ref{Tab1} show the test results with $\alpha=0.05$ for our real dataset, where ``1'' indicates that the null hypothesis ${\rm{H}}_{0k}$ was rejected. In addition, the seventh and eighth columns show the associated p-values. From the fifth column of Table \ref{Tab1}, which had many ``1'', we can see that for negative returns, many dummy variables for year, month, and day of the week cannot be ignored as covariates for predicting the EVI. The sixth column of Table \ref{Tab1} implies that for positive returns, the effects of days of the week may be meaningless in our model (\ref{Eq4.3.1}). For both negative and positive returns, the dummy variables for year may be particularly important as covariates because their p-values were quite small.
	
	Third, we predict the random effects $U_j,\ j=1, 2, \ldots, 413$ in (\ref{Eq4.3.1}). Since citing all $\tilde{u}_j$ for $j=1, 2, \ldots, 413$ is too voluminous, we only present the results for the major cryptocurrencies studied in Gkillas and Katsiampa (2018), namely Bitcoin, Ethereum, Ripple, Litecoin and Bitcoin Cash. Table \ref{Tab2} shows the predicted random effects $\tilde{u}_j$ for these five cryptocurrencies. Based on the model (\ref{Eq4.3.1}), the results in Table \ref{Tab2} mean that Bitcoin Cash had the largest EVI among the five stocks for both negative and positive returns. Thus, Bitcoin Cash may have been the riskiest of the above five cryptocurrencies in the sense that its returns were the most heavily distributed.

	\begin{table}[t]
		\captionsetup{width=0.95\linewidth}
		\caption{The predicted random effects for Bitcoin, Ethereum, Ripple, Litecoin and Bitcoin Cash.}
		\label{Tab2}
		\centering
		\scalebox{0.9}{\begin{tabular}{|c|c||ccccc|}\hline
				\multicolumn{2}{|c||}{} & Bitcoin & Ethereum & Ripple (XRP) & Litecoin & Bitcoin Cash \\\hline\hline
				\multirow{2}{*}{$\tilde{u}_j$} & Negative & $-0.319$ & $-0.224$ & $-0.319$ & $-0.304$ & $0.007$ \\\cline{2-7}
				& Positive & $-0.351$ & $-0.502$ & $-0.011$ & $-0.410$ & $0.012$ \\\hline
		\end{tabular}}
	\end{table}
	
\begin{figure}[t]
		\centering
		\includegraphics[keepaspectratio, width=100mm, angle=270]{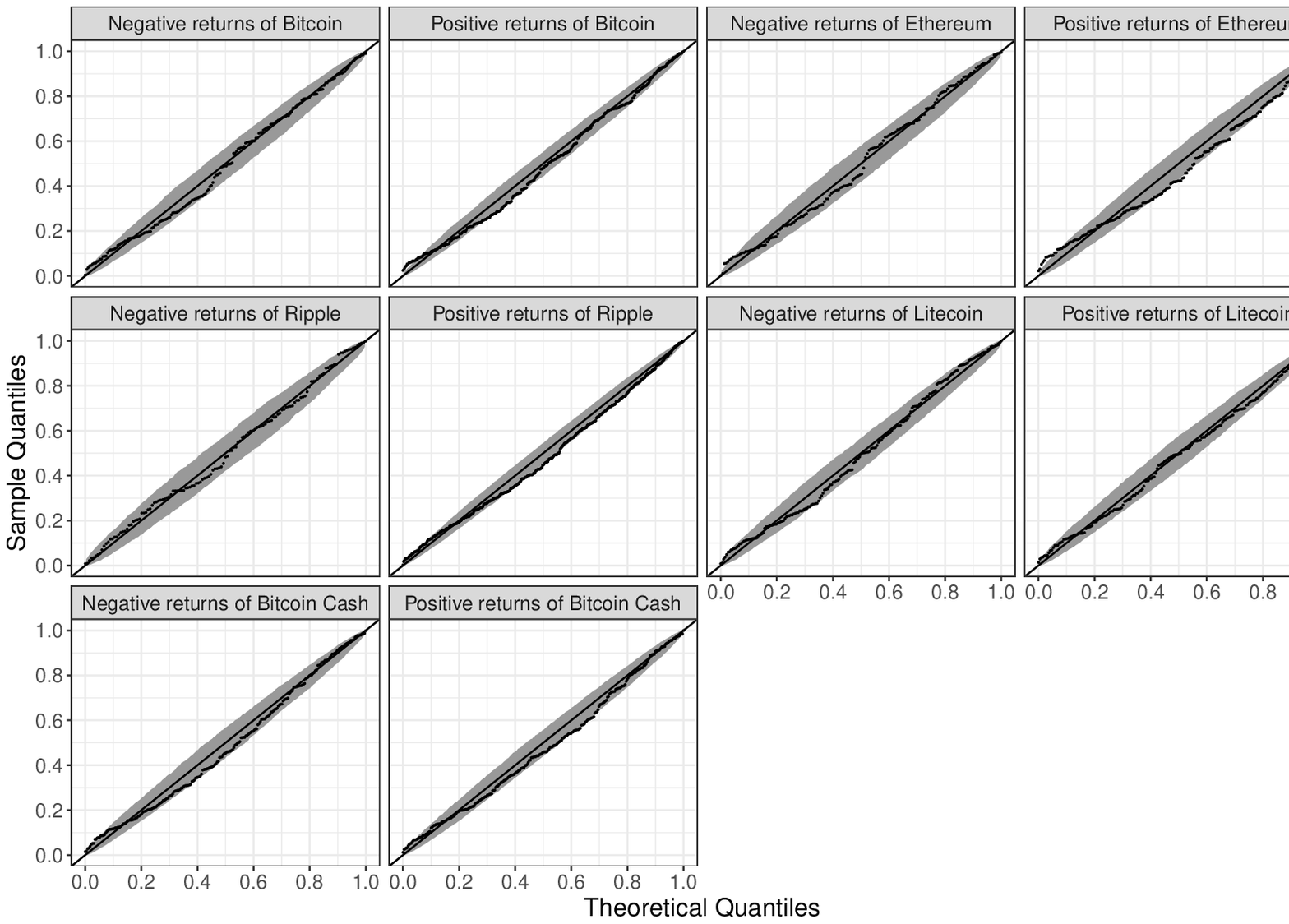}
		\caption{The Q-Q plots for Bitcoin, Ethereum, Ripple, Litecoin and Bitcoin Cash.}
		\label{Fig1}
	\end{figure}

	Finally, we evaluate the goodness of fit of the model using the similar method appeared in Wang and Tsai (2009). From (\ref{Eq2.2.2}), the distribution of $S_{ij}\coloneqq\exp[-\gamma(U_j, {\bm{X}}_{ij})^{-1}$\\
$\log(Y_{ij}/\omega_j)]$ conditional on $U_j, {\bm{X}}_{ij}$ and $Y_{ij}>\omega_j$ can be approximated by the uniform distribution on $[0, 1]$ (see, Wang and Tsai 2009). Let $\tilde{S}_{ij}$ be $S_{ij}$ with $\theta_{\rm{A}}=\hat{\theta}_{\rm{A}}$, ${\bm{\theta}}_{\rm{B}}=\hat{\bm{\theta}}_{\rm{B}}$ and $U_j=\tilde{u}_j$. Then, for each $j=1, 2, \ldots, 413$, $\mathcal{S}_j\coloneqq\{\tilde{S}_{ij}: Y_{ij}>\omega_j,\ i=1, 2, \ldots, n_j\}$ can be considered as a uniformly distributed random sample. Figure \ref{Fig1} shows the Q-Q plots of $\mathcal{S}_j$ against equally divided points on $[0, 1]$ for the five representative cryptocurrencies discussed above, where the bands are the 95\% pointwise confidence bands constructed by the function \texttt{geom\_qq\_band()} with \texttt{bandType="boot"} in the package \textsf{qqplotr} (see, Almeida, Loy, and Hofmann 2018) within \textsf{R}. In each panel of Figure \ref{Fig1}, the points were approximately aligned on a straight line with an intercept of 0 and slope of 1. This means that our model fits both negative and positive returns well for each of the five stocks.
	
	\subsection{Comparison}\label{Sec4.4}
	
\begin{figure}[t]
		\centering
		\includegraphics[keepaspectratio, width=65mm, angle=270]{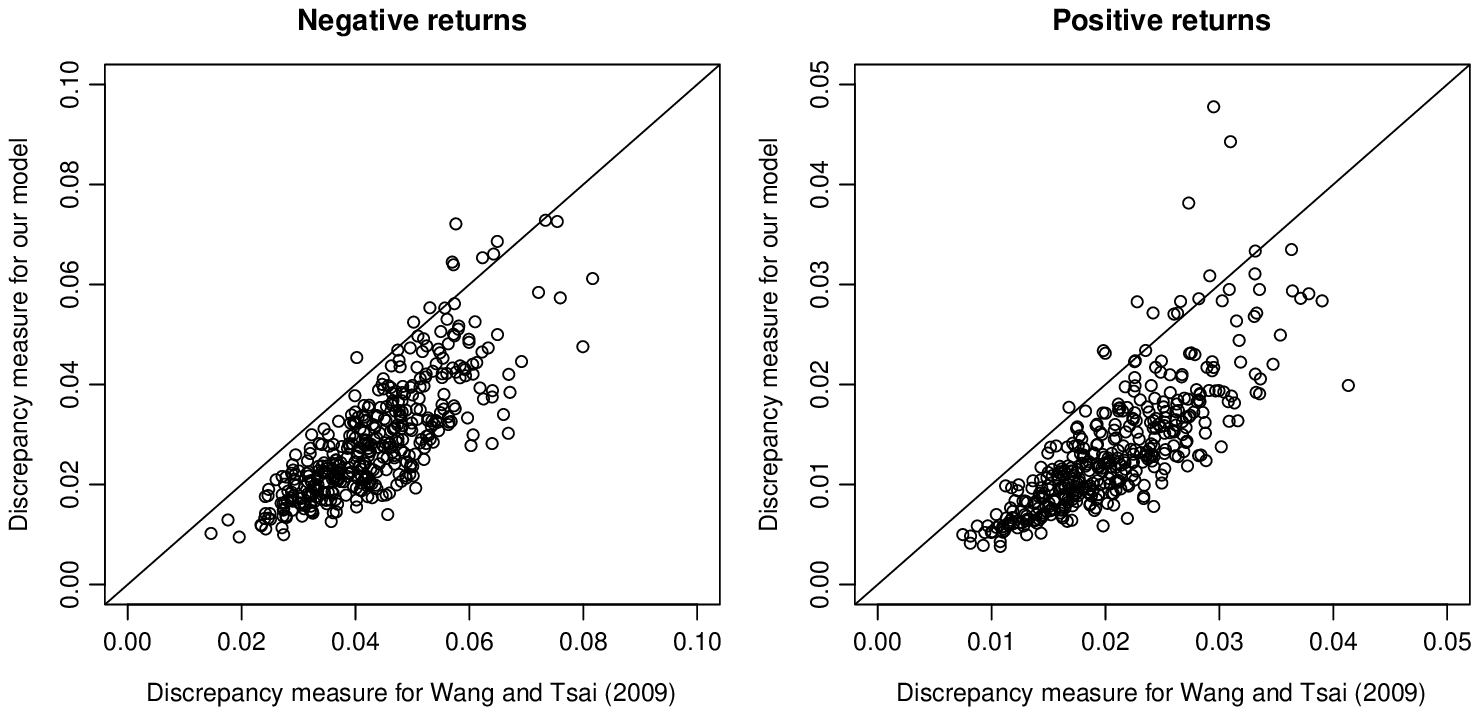}
		\caption{The scatter plot of the discrepancy measures between the model proposed by Wang and Tsai (2009) and our model, for the 413 stocks.}
		\label{Fig2}
	\end{figure}

	In this section, we compare our model (\ref{Eq4.3.1}) with the method proposed by Wang and Tsai (2009). The latter competitor applies the EVI regression to data stock by stock. To compare these two methods, we iteratively compute the following 2-fold cross-validation criterion. At each iteration step, the returns for each stock are randomly divided into training and test data. As a criterion, we adapt the discrepancy measure proposed by Wang and Tsai (2009). For each of our model and the model of Wang and Tsai (2009), the discrepancy measure for each stock is computed from the test data using the estimated parameters, predicted random effects, and selected thresholds from the training data. Note that our model and the model proposed by Wang and Tsai (2009) use the same threshold exceedances (see, Section \ref{Sec2.4}). After 100 iterations, we take the average of the discrepancy measures obtained for each stock. Each panel of Figure \ref{Fig2} shows the scatter plot of averaged discrepancy measures between our model and the model of Wang and Tsai (2009) for the 413 stocks. We can see from the results that the discrepancy measures of our model are smaller than those of Wang and Tsai (2009) for many stocks for both negative and positive returns, suggesting that our model provides better performance.
	
	\section{Discussion}\label{Sec5}
	
	In this study, we investigated the MEM for the EVI in the Pareto-type distribution for unit-level data. In other words, this study incorporated the method of SAE into EVT. As explained in Section \ref{Sec2}, the parameters of the proposed model were estimated by the maximum likelihood method, and the random effects were predicted by the conditional mode. In Section \ref{Sec3}, we established the asymptotic normality of the estimator. Together with the simulation studies in Section 1 of our supplementary material and real data example in Section \ref{Sec4}, we can conclude the following advantages of using the MEM for EVI regression. First, in extreme value analysis, the sample size is generally small for some areas because of the peak over threshold. However, as described in Sections 1.3 of our supplementary material and Section \ref{Sec4}, the common parametric part in the MEM can adequately guide the differences between areas. Interestingly, the ``borrowing of strength'' of the MEM is effective for EVI regression because it improves the bias and variance of the peak-over-threshold method (see, Section 1.3 of our supplementary material). Thus, the proposed model provides a significantly efficient tool that is an alternative to direct estimates from each area. Second, the proposed model is effective even when the number of areas is large. This is shown theoretically in Theorem \ref{Thm1} of Section \ref{Sec3.2}, while Section 1.2 of our supplementary material proves this property numerically. Furthermore, as a result supporting the use of the proposed model, in Section 1.3 of our supplementary material, the extreme value analysis using the MEM provided more reasonable results than the fully parametric model. Finally, in extreme value analysis, general EVI estimators sometimes have a large bias resulting from the approximation of the peak over threshold. However, from Theorem \ref{Thm1}, we found that when $J$ is large, the proposed estimator can be designed to reduce the bias while maintaining its stable variance. This is a somewhat surprising result, because a large number of areas typically leads to poor performance of the estimator in the fully parametric model. Thus, the MEM may be one of the effective approaches to overcome the severe problem of bias in extreme value analysis.
	
	We describe future research using the MEM for extreme value analysis. The first work of interest is the development of models such as the simultaneous autoregressive model and the conditional autoregressive model to explain the dependencies between areas (see, Rao and Molina 2015 and references therein). Such models may help provide better estimates in some applications, including spatial data. Second, it may be feasible to extend the MEM to other EVT models such as the generalized extreme value distribution and GPD. In this study, the methods derived from these models and their theoretical results were not clarified and thus require future detailed study. Third, we expect to extend the MEM to extreme quantile regression (see, Wang, Li, and He 2012; Wang and Li 2013). Finally, although this paper studied the MEM with Gaussian random effects, it may be important to consider other distributions of the random effects (see, Section 9.2 of Wu 2009 and Yavuz and Arslan 2018). The development of the MEM with non-Gaussian random effects is also an interesting future work in EVT.
	
	\section*{Acknowledgments}
	
	We would like to thank Editage (\url{https://www.editage.jp/}) for the English language editing.
	
	\section*{Data availability statement}
	
	The dataset analyzed during this study was processed from those obtained from CoinMarketCap (see, \url{https://coinmarketcap.com/}). This dataset and the associated code of \textsf{R} (https://www.r-project.org/) are available from the corresponding author on reasonable request.

	\newpage
	
	\setcounter{section}{0}

	\begin{center}
		{\Large{
				Supplemental online material for\\
				``Mixed effects models for extreme value index regression''}}
	\end{center}
	
	\bigskip
	
	\bigskip
	
	\bigskip
	
	This supplementary material supports our main article entitled ``Mixed effects models for extreme value index regression'' and is organized as follows. Section \ref{SMSec2} provides some Monte Carlo simulation studies to verify the finite sample performance of the model proposed in Section 2 of our main article. Section \ref{SMSec3} describes the technical details of the proof of Theorem 1 in Section 3.2 of our main article. Note that we use many of the symbols defined in our main article.
	
	\section{Simulation}\label{SMSec2}
	
	From Eq. (9) of our main article, we can approximate the distribution of $\log(Y_{ij}/\omega_j)$ conditional on ${\bm{U}}_j$, ${\bm{X}}_{ij}$ and $Y_{ij}>\omega_j$ by the exponential distribution, which belongs to the gamma distribution (see, Wang and Tsai 2009). Therefore, our estimator $(\hat{\bm{\theta}}_{\rm{A}}, \hat{\bm{\theta}}_{\rm{B}}, \hat{\bm{\Sigma}})$ and predictor $\tilde{\bm{u}}_j$ proposed in Section 2 of our main article can be easily implemented by using the function \texttt{glmer()} with \texttt{family=Gamma(link="log")} in the package \textsf{lme4} (see, Bates et al. 2015) within the \textsf{R} computing environment (see, R Core Team 2021). In the function \texttt{glmer()}, when $p_{\rm{A}}=1$, the integral in the log-likelihood defined in Eq. (10) of our main article is approximated by the adaptive Gauss-Hermite quadrature, and $\hat{\bm{\Sigma}}$ and $(\hat{\bm{\theta}}_{\rm{A}}, \hat{\bm{\theta}}_{\rm{B}})$ are then optimized by ``bobyqa'' and ``Nelder Mead'', respectively.
	
	In the following Sections \ref{SMSec2.1}-\ref{SMSec2.3}, we investigate the performance of our estimator $(\hat{\bm{\theta}}_{\rm{A}}, \hat{\bm{\theta}}_{\rm{B}}, \hat{\bm{\Sigma}})$ and predictor $\tilde{\bm{u}}_j$ through some simulation studies using the above package.
	
	\subsection{Practicality of asymptotic normality of the estimator}\label{SMSec2.1}
	
	In this section, we illustrate the applicability of our asymptotic normality constructed in Corollary 1 of our main article to finite samples. This is positioned as a preliminary study for hypothesis testing on a real data example in Section 4 of our main article. 
	
	We simulate the dataset $\{(Y_{ij}, {\bm{X}}_{ij})\}_{i\in\mathcal{N}_j, j\in\mathcal{J}}$ as follows. Let denote ${\bm{X}}_{ij}=(X_{ij}^{(1)}, X_{ij}^{(2)})^\top\in\mathbb{R}^2$ and set $X_{ij}^{(1)}\equiv1$ for $i\in\mathcal{N}_j$ and $j\in\mathcal{J}$. First, we independently generate $\{X_{ij}^{(2)}\}_{i\in\mathcal{N}_j, j\in\mathcal{J}}$ from the standard normal distribution or uniform distribution on $[-\sqrt{3}, \sqrt{3}]$. Note that in both covariate cases, $X_{ij}^{(2)}$ has zero mean and unit variance. In the next step, we obtain an independent sample $\{U_j\}_{j\in\mathcal{J}}$ from $N(0, \sigma^2)$ with $\sigma^2=0.2$. Finally, for each $i\in\mathcal{N}_j$ and $j\in\mathcal{J}$, we generate $Y_{ij}$ using a given conditional response distribution $F(\cdot\mid U_j, X_{ij}^{(2)})$. The same data generation procedure will be used in Section \ref{SMSec2.2}. Here, to obtain $\{Y_{ij}\}_{i\in\mathcal{N}_j, j\in\mathcal{J}}$, we use the Pareto distribution 
	\begin{equation}
		F(y\mid U_j, X_{ij}^{(2)})=1-y^{-1/\gamma(U_j, X_{ij}^{(2)})}\label{SMEq2.1.1}
	\end{equation}
	and apply the nested error regression type model formulated as Eq. (7) of our main article with ${\bm{X}}_{{\rm{A}}ij}=X_{ij}^{(1)}$ and ${\bm{X}}_{{\rm{B}}ij}=X_{ij}^{(2)}$, i.e.,
	\begin{equation}
		\gamma(U_j, X_{ij}^{(2)})=\exp\left(\theta_{\rm{A}}^{(1)}+U_j+\theta_{\rm{B}}^{(2)}X_{ij}^{(2)}\right),\label{SMEq2.1.2}
	\end{equation}
	where $(\theta_{\rm{A}}^{(1)}, \theta_{\rm{B}}^{(2)})=(-0.5, 0.2)$. Let us denote the proposed estimator of $(\theta_{\rm{A}}^{(1)}, \theta_{\rm{B}}^{(2)}, \sigma^2)$ by $(\hat{\theta}_{\rm{A}}^{(1)}, \hat{\theta}_{\rm{B}}^{(2)}, \hat{\sigma}^2)$. Because the Pareto distribution (\ref{SMEq2.1.1}) satisfies Eq. (12) of our main article with $\beta(U_j, X_{ij}^{(2)})=\infty$, the estimator does not have the asymptotic bias described in Remark 1 in Section 3.2 of our main article. Therefore, we do not use the thresholds $\omega_j,\ j\in\mathcal{J}$, and thus the effective sample size $n_{j0}$ for each $j\in\mathcal{J}$ is unchanged from $n_j$.
	
	\begin{figure}
		\centering
		\includegraphics[keepaspectratio, width=140mm]{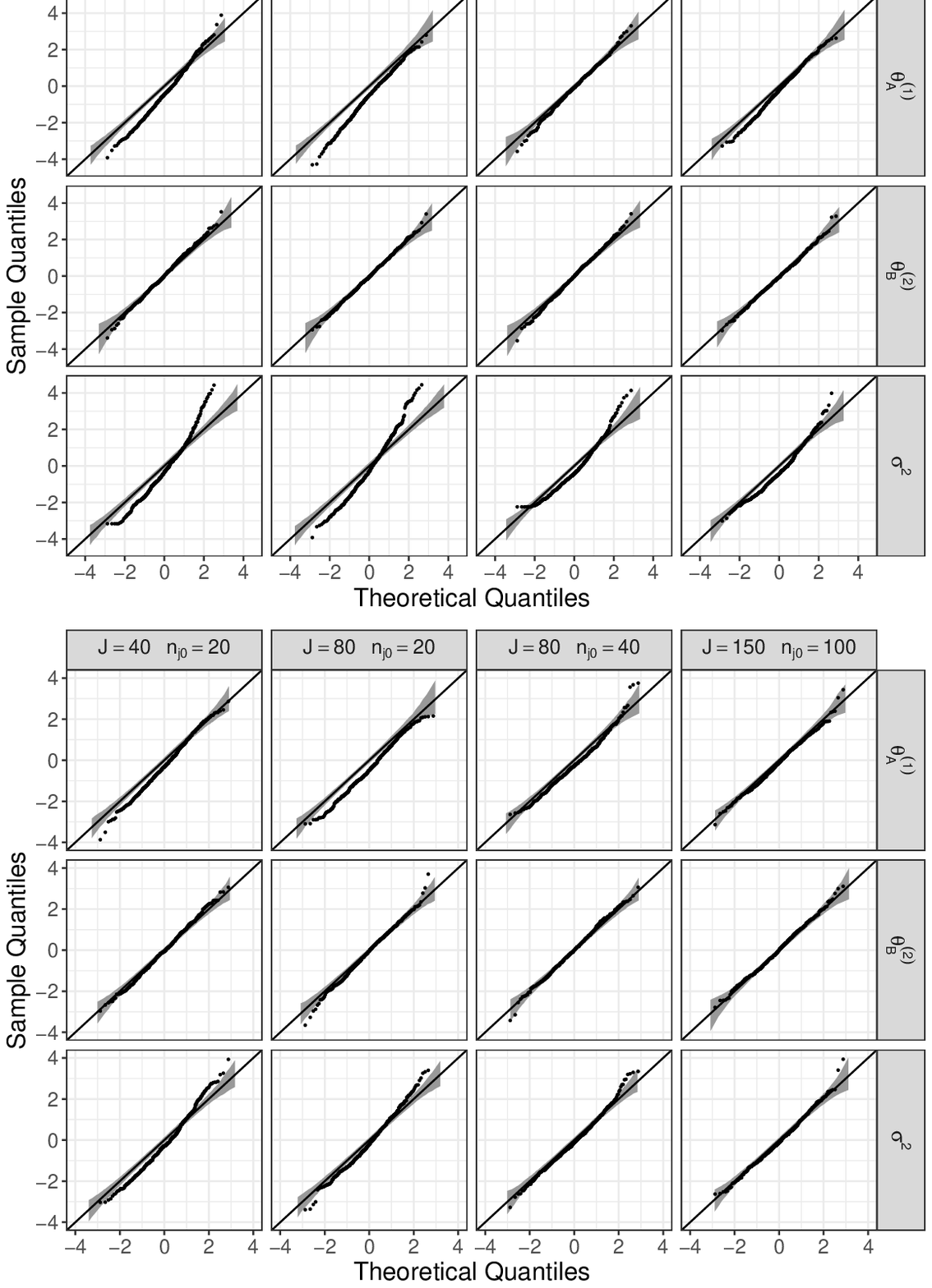}
		\caption{Results of the simulation with the normal covariate: Q-Q plots for the standardized estimates against $N(0, 1)$.}
		\label{SMFig2.1.1}
	\end{figure}
	
	\begin{figure}
		\centering
		\includegraphics[keepaspectratio, width=140mm]{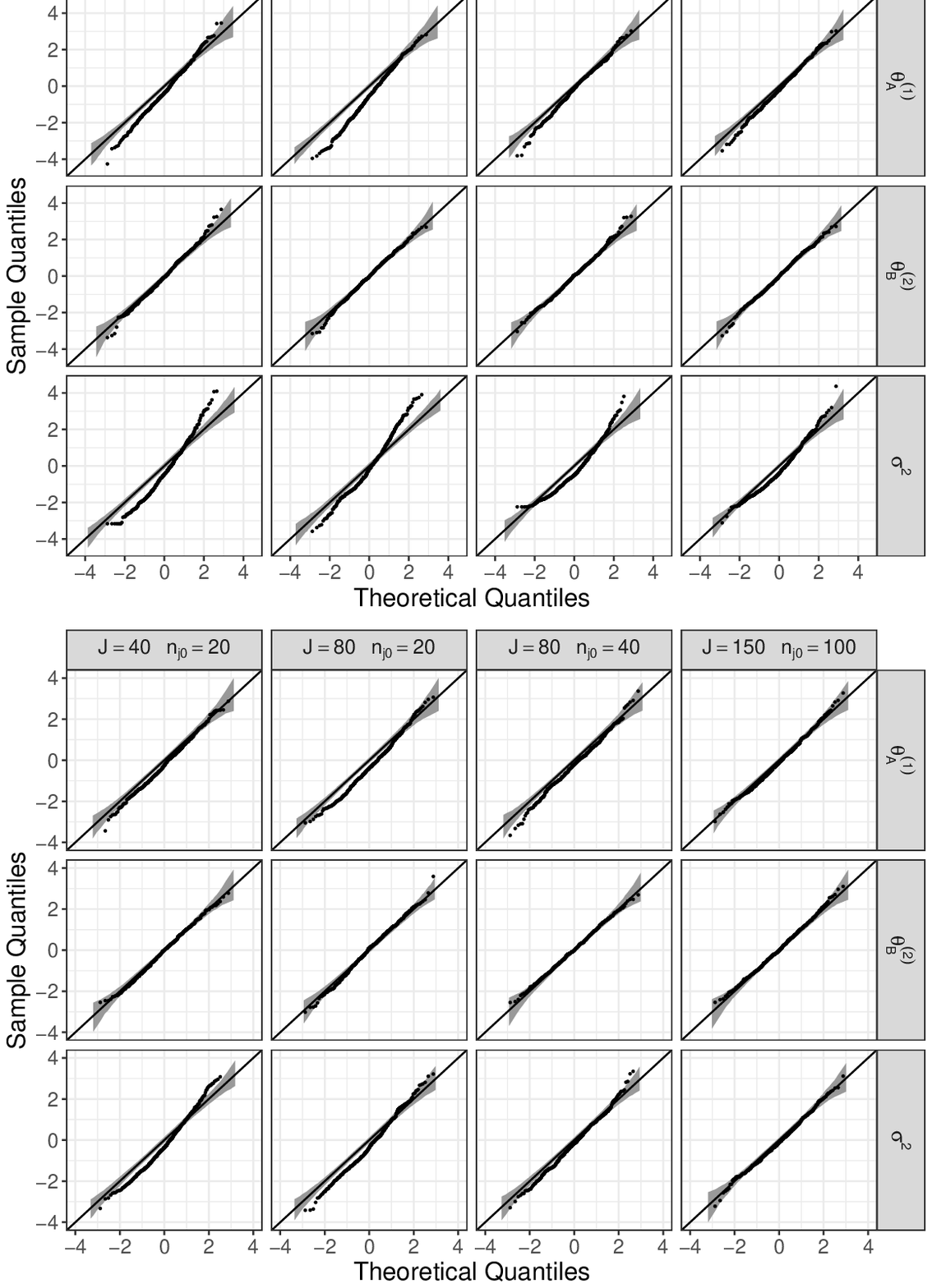}
		\caption{Results of the simulation with the uniform covariate: Q-Q plots for the standardized estimates against $N(0, 1)$.}
		\label{SMFig2.1.2}
	\end{figure}
	
	Under the above model setups, from Corollary 1 of our main article, $J^{1/2}(\hat{\theta}_{\rm{A}}^{(1)}-\theta_{\rm{A}}^{(1)})/\sigma$, $(Jn_0)^{1/2}(\hat{\theta}_{\rm{B}}^{(2)}-\theta_{\rm{B}}^{(2)})$ and $J^{1/2}(\hat{\sigma}^2-\sigma^2)/(\sqrt{2}\sigma^2)$ are asymptotically distributed as $N(0, 1)$. Note that this simulation setting satisfies ${\bm{\Omega}}_{\rm{B}}=1$ in Corollary 1 of our main article. To obtain the empirical distributions of the above standardized estimators, we use 500 datasets and repeatedly estimate the parameters $\{\theta_{\rm{A}}^{(1)}, \theta_{\rm{B}}^{(2)}, \sigma^2\}$ from each dataset. Figures \ref{SMFig2.1.1} and \ref{SMFig2.1.2} show the Q-Q plots for the obtained standardized estimates against $N(0, 1)$ for the normal covariate and uniform covariate, respectively. In these figures, $(J, n_{j0})$ varies by column as $(20, 10)$, $(40, 10)$, $(10, 20)$, $(20, 20)$, $(40, 20)$, $(80, 20)$, $(80, 40)$, and $(150, 100)$. Furthermore, the bands in each panel are the 95\% pointwise confidence bands constructed by the function \texttt{geom\_qq\_band()} with \texttt{bandType="boot"} in the package \textsf{qqplotr} (see, Almeida, Loy, and Hofmann 2018) within \textsf{R}. If all generated $U_1, U_2, \ldots, U_J$ are close to each other, $\sigma^2$ may be estimated to be zero by \texttt{glmer()}. Thus, for $J=10$ and $J=20$, the Q-Q plot contained several equal values for $\sigma^2$. Comparing Figures \ref{SMFig2.1.1} and \ref{SMFig2.1.2}, the type of the distribution of ${X}_{ij}^{(2)}$ did not significantly affect the results. We can see from Figures \ref{SMFig2.1.1} and \ref{SMFig2.1.2} that for $\theta_{\rm{A}}^{(1)}$ and $\sigma^2$, the empirical distribution of the standardized estimators had heavier tails than $N(0, 1)$, but this tendency disappeared with increasing $J$ and $n_{j0}$. In contrast to $\theta_{\rm{A}}^{(1)}$ and $\sigma^2$, the Q-Q plot for $\theta_{\rm{B}}^{(2)}$ was good for all pairs of $(J, n_{j0})$. Consequently, these results reflect the claims of Corollary 1 of our main article.
	
	\subsection{Behavior of the estimator for numerous areas}\label{SMSec2.2}
	
	In this section, we examine the numerical performance of the estimator for large $J$. To obtain the dataset $\{(Y_{ij}, {\bm{X}}_{ij})\}_{i\in\mathcal{N}_j, j\in\mathcal{J}}$ according to the procedure in Section \ref{SMSec2.1}, we use (\ref{SMEq2.1.2}) and the following conditional distribution (a) or (b) instead of (\ref{SMEq2.1.1}):
	
	\renewcommand{\theenumi}{\alph{enumi}}
	\renewcommand{\labelenumi}{(\theenumi)}
	\begin{enumerate}
		\item Student's $t$-distribution
		\begin{align*}
			&F(y\mid U_j, X_{ij}^{(2)})\\
			&=\int_{-\infty}^y\frac{\Gamma([\nu(U_j, X_{ij}^{(2)})+1]/2)}{\sqrt{\nu(U_j, X_{ij}^{(2)})\pi}\Gamma(\nu(U_j, X_{ij}^{(2)})/2)}\left[1+\frac{t^2}{\nu(U_j, X_{ij}^{(2)})}\right]^{-\left[\nu(U_j, X_{ij}^{(2)})+1\right]/2}dt
		\end{align*}
		with $\nu(U_j, X_{ij}^{(2)})\coloneqq\gamma(U_j, X_{ij}^{(2)})^{-1}$, where $\Gamma(\cdot)$ is a gamma function. For Eq. (12) of our main article, this distribution belongs to the Pareto-type distribution with $\beta(U_j, X_{ij}^{(2)})\equiv2$.
		\item Burr distribution
		\begin{equation*}
			F(y\mid U_j, X_{ij}^{(2)})=1-\left[\frac{\eta}{\eta+y^{\tau(U_j, X_{ij}^{(2)})}}\right]^\kappa
		\end{equation*}
		with $\eta=1$, $\kappa=1$ and $\tau(U_j, X_{ij}^{(2)})\coloneqq\gamma(U_j, X_{ij}^{(2)})^{-1}$. The Burr distribution satisfies Eq. (12) of our main article with $\beta(U_j, X_{ij}^{(2)})=\tau(U_j, X_{ij}^{(2)})$.
	\end{enumerate}
	For (a), we obtain a sample directly from the $t$-distribution for a given non-integer degree of freedom. As described in Remark 1 of our main article, the proposed estimator is more biased for smaller $\beta(U_j, X_{ij}^{(2)})$.
	
	\begin{figure}
		\centering
		\includegraphics[keepaspectratio, width=140mm]{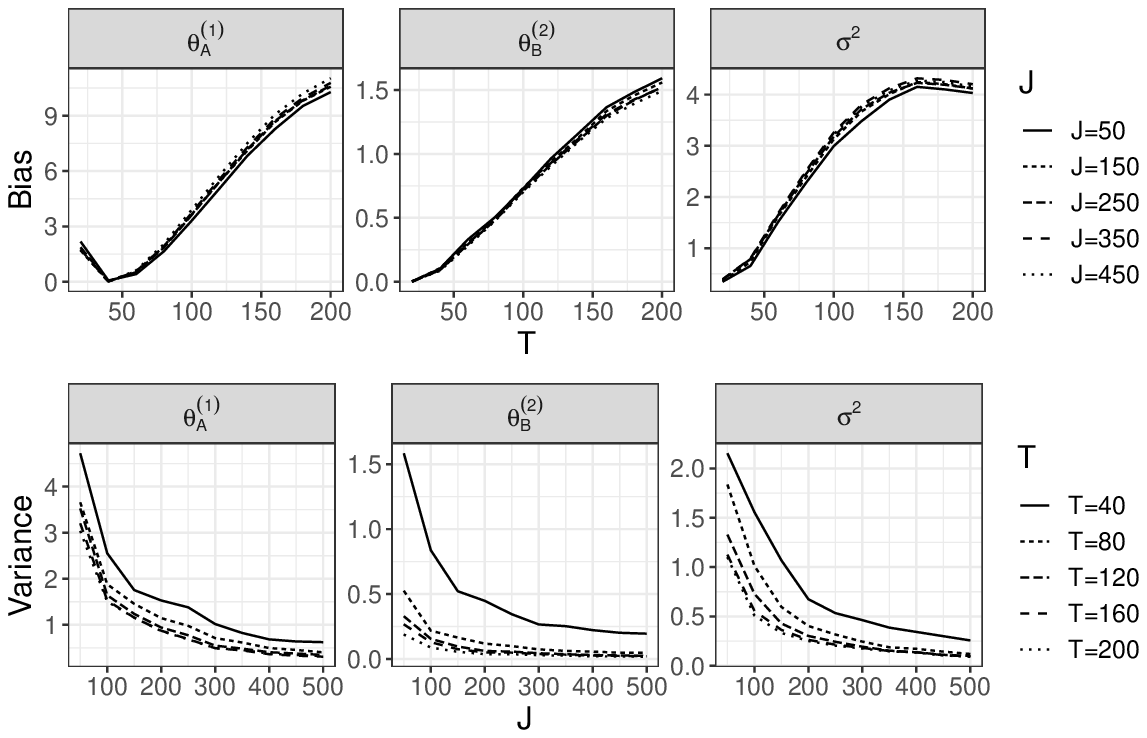}
		\caption{Simulation results for the distribution (a) with the normal covariate: Variations of the sample squared bias $(\times10^3)$ and variance $(\times10^3)$ of the estimator with respect to $J$ and $T$.}
		\label{SMFig2.2.1}
	\end{figure}
	
	\begin{figure}
		\centering
		\includegraphics[keepaspectratio, width=140mm]{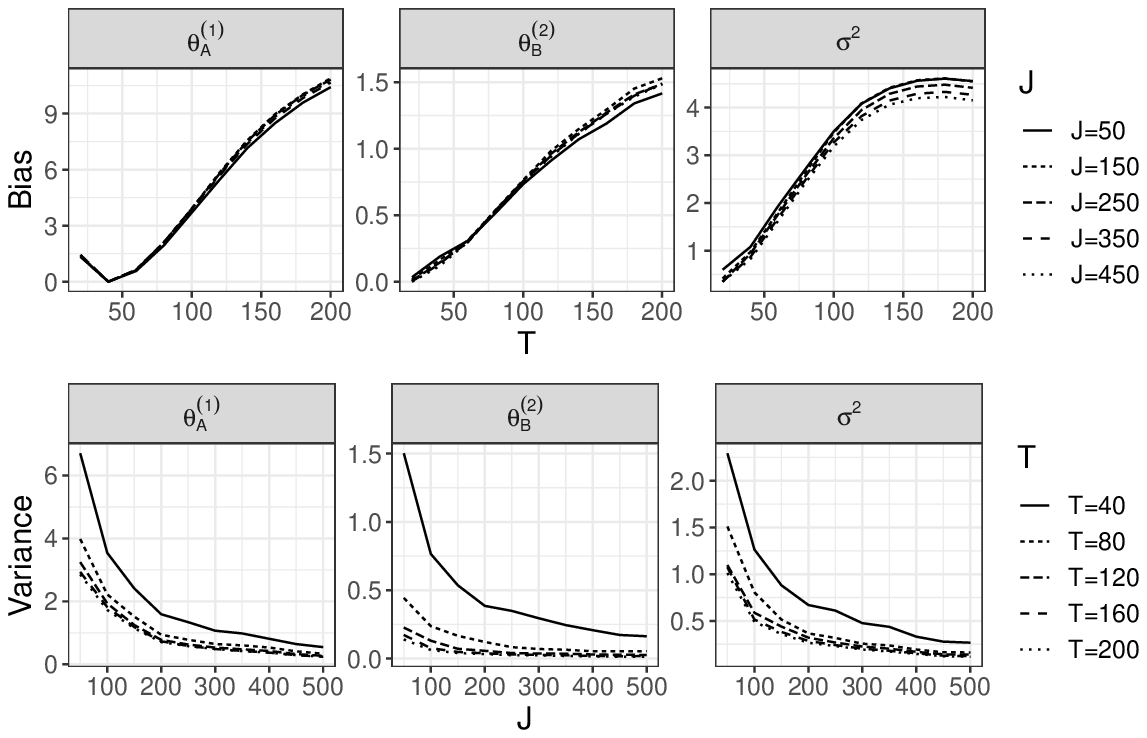}
		\caption{Simulation results for the distribution (a) with the uniform covariate: Variations of the sample squared bias $(\times10^3)$ and variance $(\times10^3)$ of the estimator with respect to $J$ and $T$.}
		\label{SMFig2.2.2}
	\end{figure}
	
	\begin{figure}
		\centering
		\includegraphics[keepaspectratio, width=140mm]{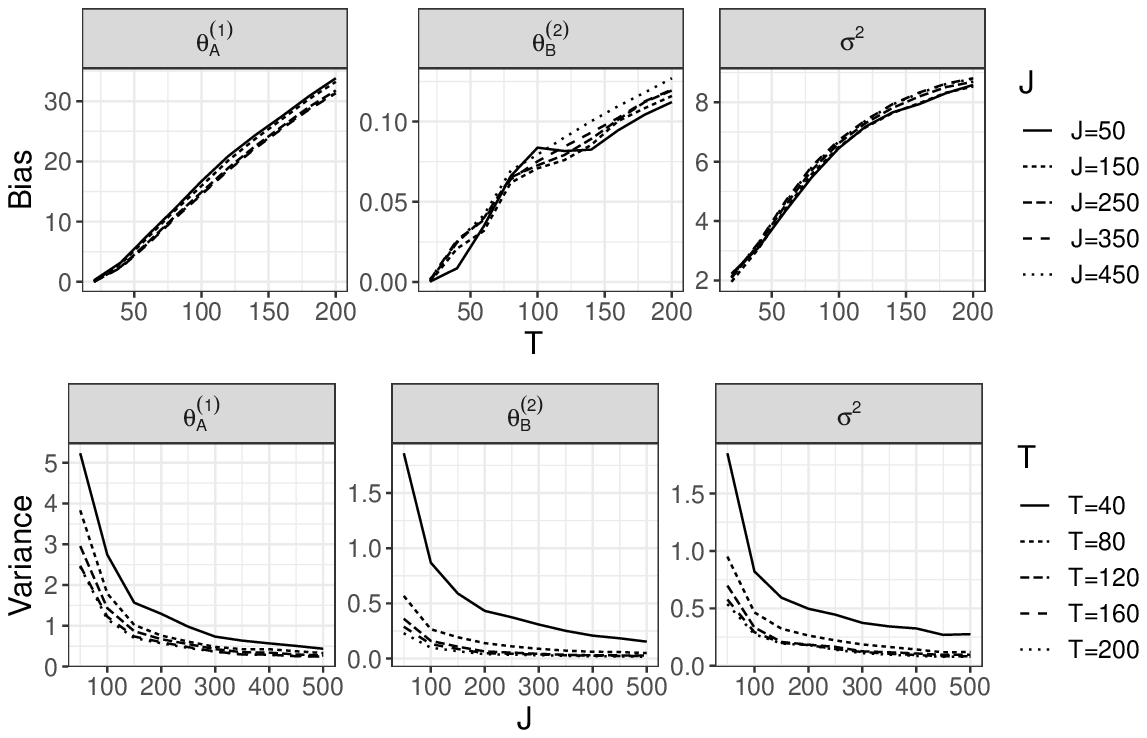}
		\caption{Simulation results for the distribution (b) with the normal covariate: Variations of the sample squared bias $(\times10^3)$ and variance $(\times10^3)$ of the estimator with respect to $J$ and $T$.}
		\label{SMFig2.2.3}
	\end{figure}
	
	\begin{figure}
		\centering
		\includegraphics[keepaspectratio, width=140mm]{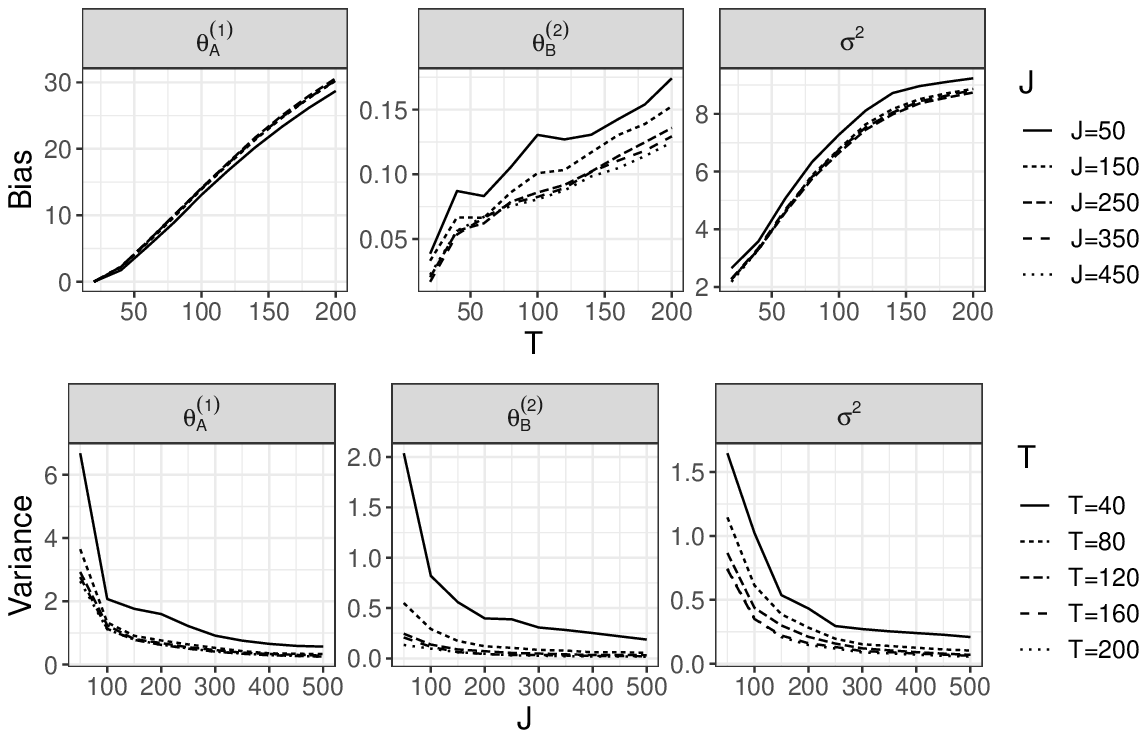}
		\caption{Simulation results for the distribution (b) with the uniform covariate: Variations of the sample squared bias $(\times10^3)$ and variance $(\times10^3)$ of the estimator with respect to $J$ and $T$.}
		\label{SMFig2.2.4}
	\end{figure}
	
	In advance, we generate the unit-level data with 500 areas and $n_j=1000$ using the above procedure. Then, we use a part of the dataset according to the following rules. The number of areas $J$ is increased by a factor of 10 from 50 to 500. Furthermore, for the discrepancy measure described in Section 2.4 of our main article, we use the 10th to $T$th largest responses in the $j$th area as the candidates for the $j$th threshold $\omega_j$, where $T$ varies from 20 to 200 in increments of 20. Roughly speaking, a smaller $T$ means that higher thresholds $\omega_j,\ j\in\mathcal{J}$ are chosen. For each $J$ and $T$, we obtain the estimates $\{\hat{\theta}_{\rm{A}}^{(1)}, \hat{\theta}_{\rm{B}}^{(2)}, \hat{\sigma}^2\}$. Following the above rules, we iterate the estimation for 100 sets of unit-level data. Figures \ref{SMFig2.2.1}-\ref{SMFig2.2.4} show the sample squared bias and variance of the estimator for each $J$ and $T$. The details can be found in the description of each figure. Overall, we can see that our estimator remained stable when $J$ was sufficiently large, even when $T$ was small. Based on the first row of each figure, our estimator then did not suffer from a large bias for large $J$. Such results guarantee the considerations in Remark 2 of Section 3.2 of our main article. As described in Remark 1 of our main article, from the second row of each figure, the variance of $\hat{\theta}_{\rm{B}}^{(2)}$ was strongly dependent on both $J$ and $T$ when $J$ was small, while the variances of $\hat{\theta}_{\rm{A}}^{(1)}$ and $\hat{\sigma}^2$ were almost unaffected by $T$.
	
	\subsection{Borrowing of strength for extremes}\label{SMSec2.3}
	
	In this section, we compare our model, the fully parametric model defined in Eqs. (2) and (3) of our main article, and model proposed by Wang and Tsai (2009). The use of the model proposed by Wang and Tsai (2009) implies that areas are analyzed separately. To obtain the estimates for the latter two parametric models, we use the function \texttt{glm()} with \texttt{family=Gamma(link="log")} within \textsf{R}.
	
	Unlike Sections \ref{SMSec2.1} and \ref{SMSec2.2}, the dataset $\{(Y_{ij}, {\bm{X}}_{ij})\}_{i\in\mathcal{N}_j, j\in\mathcal{J}}$ is simulated from the fully parametric model as follows. Let $J$ be divided into $J=J^\dag+J^\ddag$ and set $J^\dag=130$, $J^\ddag=20$, $n_1=\cdots=n_{J^\dag}=100$ and $n_{J^\dag+1}=\cdots=n_{J^\dag+J^\ddag}=20$. We generate $\{X_{ij}^{(2)}\}_{i\in\mathcal{N}_j, j\in\mathcal{J}}$ from $N(0, 1)$. To generate $\{Y_{ij}\}_{i\in\mathcal{N}_j, j\in\mathcal{J}}$, we then combine the two distributions 
	\begin{equation}
		F_j^\dag(y\mid X_{ij}^{(2)})=1-y^{-1/\gamma_j(X_{ij}^{(2)})},\quad j=1, 2, \ldots, J^\dag\label{SMEq2.3.1}
	\end{equation}
	and
	\begin{equation}
		F_j^\ddag(y\mid X_{ij}^{(2)})=1-\frac{1.5y^{-1/\gamma_j(X_{ij}^{(2)})}}{1+0.5y^{-1/\gamma_j(X_{ij}^{(2)})}},\quad j=J^\dag+1, J^\dag+2, \ldots, J^\dag+J^\ddag,\label{SMEq2.3.2}
	\end{equation}
	where $F_j^\ddag$ is described in detail in Section 4.1 of Wang and Tsai (2009). Then, the extreme value index $\gamma_j(X_{ij}^{(2)})$ is modeled as $\gamma_j(X_{ij}^{(2)})=\exp(\theta_{j{\rm{A}}}^{(1)}+\theta_{\rm{B}}^{(2)}X_{ij}^{(2)})$, where $\theta_{\rm{B}}^{(2)}=0.2$ and the parameters $\theta_{j{\rm{A}}}^{(1)},\ j\in\mathcal{J}$ are assigned the following two types. One $\theta_{j{\rm{A}}}^{(1)}$ is defined as the $j/(J^\dag+1)$-th quantile of $N(0, 1/12)$ for $j=1, 2, \ldots, J^\dag$ and $(j-J^\dag)/(J^\ddag+1)$-th quantile of $N(0, 1/12)$ for $j=J^\dag+1, J^\dag+2, \ldots, J^\dag+J^\ddag$, and the other $\theta_{j{\rm{A}}}^{(1)}$ is fixed as $-0.5+0.5(j-1)/(J^\dag-1)$ for $j=1, 2, \ldots, J^\dag$ and $-0.5+0.5(j-J^\dag-1)/(J^\ddag-1)$ for $j=J^\dag+1, J^\dag+2, \ldots, J^\dag+J^\ddag$. These two settings of $\theta_{j{\rm{A}}}^{(1)}$ are denoted by Normal type and Uniform type, respectively.
	
	\begin{table}[t]
		\captionsetup{width=0.95\linewidth}
		\caption{Sample bias and variance of the estimator for the three models.}
		\label{SMTab2.3.1}
		\centering
		\begin{tabular}{|l|l||rrr|}\hline
			\multicolumn{2}{|c|}{} & Our model & Fully parametric model & Direct estimates \\\hline\hline
			\multicolumn{2}{|c}{Normal type} & & & \\\hline
			\multirow{4}{*}{$\theta_{j{\rm{A}}}^{(1)}$} & $\text{Bias}^\dag$ & $-1.77\times10^{-3}$ & $-4.39\times10^{-3}$ & $-9.50\times10^{-3}$ \\
			& $\text{Bias}^\ddag$ & $1.20\times10^{-1}$ & $1.77\times10^{-1}$ & $1.57\times10^{-1}$ \\
			& $\text{Variance}^\dag$ & $7.89\times10^{-3}$ & $9.97\times10^{-3}$ & $1.01\times10^{-2}$ \\
			& $\text{Variance}^\ddag$\ & $1.69\times10^{-2}$ & $4.25\times10^{-2}$ & $4.59\times10^{-2}$ \\\hline
			\multirow{2}{*}{$\theta_{\rm{B}}^{(2)}$} & $\text{Bias}$ & $2.00\times10^{-1}$ & $2.00\times10^{-1}$ & $2.00\times10^{-1}$ \\
			& $\text{Variance}$ & $6.78\times10^{-5}$ & $6.78\times10^{-5}$ & $1.23\times10^{-4}$ \\\hline\hline
			\multicolumn{2}{|c}{Uniform type} & & & \\\hline
			\multirow{4}{*}{$\theta_{j{\rm{A}}}^{(1)}$} & $\text{Bias}^\dag$ & $-3.07\times10^{-3}$ & $-5.56\times10^{-3}$ & $-1.06\times10^{-2}$ \\
			& $\text{Bias}^\ddag$ & $1.26\times10^{-1}$ & $1.80\times10^{-1}$ & $1.60\times10^{-1}$ \\
			& $\text{Variance}^\dag$ & $8.07\times10^{-3}$ & $9.98\times10^{-3}$ & $1.01\times10^{-2}$ \\
			& $\text{Variance}^\ddag$\ & $1.73\times10^{-2}$ & $4.10\times10^{-2}$ & $4.50\times10^{-2}$ \\\hline
			\multirow{2}{*}{$\theta_{\rm{B}}^{(2)}$} & $\text{Bias}$ & $2.00\times10^{-1}$ & $2.00\times10^{-1}$ & $1.99\times10^{-1}$ \\
			& $\text{Variance}$ & $7.32\times10^{-5}$ & $7.34\times10^{-5}$ & $1.13\times10^{-4}$ \\\hline
		\end{tabular}
	\end{table}
	
	Without introducing thresholds, we estimate $\theta_{j{\rm{A}}}^{(1)},\ j\in\mathcal{J}$ and $\theta_{\rm{B}}^{(2)}$ by each of our model, the fully parametric model expressed as Eqs. (2) and (3) of our main article, and model proposed by Wang and Tsai (2009). In our model, estimates for $\theta_{{\rm{A}}j}^{(1)},\ j\in\mathcal{J}$ are obtained by combining the estimator and predictor proposed in Section 2 of our main article. For the model proposed by Wang and Tsai (2009), different estimates of $\theta_{\rm{B}}^{(2)}$ are obtained across the areas, but the mean of these is used as the estimate of $\theta_{\rm{B}}^{(2)}$. Under the above simulation settings, if the areas were analyzed separately, the estimates for the areas from (\ref{SMEq2.3.2}) would have large bias and variance. Therefore, we are interested in whether these direct estimates can be improved by simultaneously analyzing the areas using our model. To verify this, we evaluate the sample bias and variance of the estimator for each method. Table \ref{SMTab2.3.1} shows the results based on 500 sets of unit-level data, where the superscripts $\dag$ and $\ddag$ in this table refer to the means for $j=1, 2, \ldots, J^\dag$ and for $j=J^\dag+1, J^\dag+2, \ldots, J^\dag+J^\ddag$, respectively. According to Table \ref{SMTab2.3.1}, our model produced the most stable estimates of the three models. In particular, for $\theta_{j{\rm{A}}}^{(1)},\ j=J^\dag+1, J^\dag+2, \ldots, J^\dag+J^\ddag$, the variance of the estimates was dramatically improved using our model rather than the direct estimates. In addition, we can see that in extreme value analysis, the ``borrowing of strength'' of the mixed effects model also contributes to reducing the bias of the estimator. In the fully parametric model, the bias of the estimator was larger than that of the direct estimates. Meanwhile, even when the assumption of normality of the random effects in Eq. (4) of our main article was not satisfied, our model was still favorable.
	
	\section{Proof of theorems}\label{SMSec3}
	
	We give the proof of Theorem 1 in Section 3.2 of our main article. For convenience, we introduce some new symbols:
	\begin{itemize}
		\item ${\bm{\theta}}_{\rm{C}}\coloneqq{\rm{vech}}({\bm{\Sigma}})$, ${\bm{\theta}}_{\rm{C}}^0\coloneqq{\rm{vech}}({\bm{\Sigma}}_0)$ and $\hat{\bm{\theta}}_{\rm{C}}\coloneqq{\rm{vech}}(\hat{\bm{\Sigma}})$, which are $p_{\rm{C}}\coloneqq p_{\rm{A}}(p_{\rm{A}}+1)/2$-dimensional vectors.
		\item ${\bm{\theta}}\coloneqq({\bm{\theta}}_{\rm{A}}^\top, {\bm{\theta}}_{\rm{B}}^\top, {\bm{\theta}}_{\rm{C}}^\top)^\top$, ${\bm{\theta}}^0\coloneqq(({\bm{\theta}}_{\rm{A}}^0)^\top, ({\bm{\theta}}_{\rm{B}}^0)^\top, ({\bm{\theta}}_{\rm{C}}^0)^\top)^\top$ and $\hat{\bm{\theta}}\coloneqq(\hat{\bm{\theta}}_{\rm{A}}^\top, \hat{\bm{\theta}}_{\rm{B}}^\top, \hat{\bm{\theta}}_{\rm{C}}^\top)^\top$.
		\item For each $j\in\mathcal{J}$, we denote
		\begin{align*}
			\ell_j({\bm{\theta}})&\coloneqq\log\int_{\mathbb{R}^{p_{\rm{A}}}}\phi({\bm{u}}; {\bm{0}}, {\bm{\theta}}_{\rm{C}})\exp\left(\sum_{i=1}^{n_j}\biggl\{-\left({\bm{\theta}}_{\rm{A}}+{\bm{u}}\right)^\top{\bm{X}}_{{\rm{A}}ij}-{\bm{\theta}}_{\rm{B}}^\top{\bm{X}}_{{\rm{B}}ij}\biggr.\right.\nonumber\\
			&\quad\Biggl.\left.-\exp\left[-\left({\bm{\theta}}_{\rm{A}}+{\bm{u}}\right)^\top{\bm{X}}_{{\rm{A}}ij}-{\bm{\theta}}_{\rm{B}}^\top{\bm{X}}_{{\rm{B}}ij}\right]\log\frac{Y_{ij}}{\omega_{(J, n_j)}}\right\}I(Y_{ij}>\omega_{(J, n_j)})\Biggr)d{\bm{u}},
		\end{align*}
		where $\phi({\bm{u}}; {\bm{0}}, {\bm{\theta}}_{\rm{C}})\coloneqq\phi({\bm{u}}; {\bm{0}}, {\bm{\Sigma}})$. In Eq. (10) of our main article, the approximated log-likelihood $\ell({\bm{\theta}}_{\rm{A}}, {\bm{\theta}}_{\rm{B}}, {\bm{\Sigma}})$ can be redefined as $\ell({\bm{\theta}})\coloneqq\sum_{j=1}^J\ell_j({\bm{\theta}})$.
		\item For any smooth function $R_1: \mathbb{R}^d\to\mathbb{R}; {\bm{z}}\mapsto R_1({\bm{z}})$, we denote $\nabla R_1({\bm{z}})\coloneqq(\partial/\partial{\bm{z}})R_1({\bm{z}})\in\mathbb{R}^d$ and $\nabla^2 R_1({\bm{z}})\coloneqq(\partial^2/\partial{\bm{z}}\partial{\bm{z}}^\top)R_1({\bm{z}})\in\mathbb{R}^{d\times d}$. In particular, we denote $\nabla_{{\bm{z}}_0}R_1({\bm{z}})\coloneqq(\partial/\partial{\bm{z}}_0)R_1({\bm{z}})$ and $\nabla_{{\bm{z}}_1{\bm{z}}_2}^2R_1({\bm{z}})\coloneqq(\partial^2/\partial{\bm{z}}_1\partial{\bm{z}}_2^\top)R_1({\bm{z}})$, where ${\bm{z}}_0$, ${\bm{z}}_1$ and ${\bm{z}}_2$ are part of ${\bm{z}}$. As a special case, for any smooth real-valued function $R_2({\bm{\theta}})$ of ${\bm{\theta}}=({\bm{\theta}}_{\rm{A}}^\top, {\bm{\theta}}_{\rm{B}}^\top, {\bm{\theta}}_{\rm{C}}^\top)^\top$, we simply write $\nabla_{\rm{K}} R_2({\bm{\theta}})\coloneqq\nabla_{{\bm{\theta}}_{\rm{K}}}R_2({\bm{\theta}})$ and $\nabla_{{\rm{K}}_1{\rm{K}}_2}^2R_2({\bm{\theta}})\coloneqq\nabla_{{\bm{\theta}}_{{\rm{K}}_1}{\bm{\theta}}_{{\rm{K}}_2}}^2R_2({\bm{\theta}})$ for ${\rm{K}}, {\rm{K}}_1, {\rm{K}}_2\in\{{\rm{A}}, {\rm{B}}, {\rm{C}}\}$.
		\item For any column vector ${\bm{z}}$, we denote ${\bm{z}}^{\otimes2}\coloneqq{\bm{z}}{\bm{z}}^\top$.
		\item Let ${\bm{\Upsilon}}_{(J, n_0)}$ be the $(p_{\rm{A}}+p_{\rm{B}}+p_{\rm{C}})$-diagonal matrix with ${\rm{diag}}({\bm{\Upsilon}}_{(J, n_0)})=(J^{1/2}{\bm{1}}_{p_{\rm{A}}}^\top, J^{1/2}n_0^{1/2}{\bm{1}}_{p_{\rm{B}}}^\top,$\\
		$ J^{1/2}{\bm{1}}_{p_{\rm{C}}}^\top)^\top$, where ${\bm{1}}_d$ is the $d$-dimensional vector with all elements equal to 1.
	\end{itemize}
	
	Let denote $D_j(\bm{u})\coloneqq n_jP(Y_{ij}>\omega_{(J, n_j)}\mid {\bm{U}}_j={\bm{u}})/n_0$. For each $j\in\mathcal{J}$, $D_j(\bm{u})$ conditional on ${\bm{U}}_j={\bm{u}}$ satisfies the following Lemma \ref{Lem1}.
	
	\begin{lemma}\label{Lem1}
		Suppose that (A3) and (A4) hold. Then, under given ${\bm{U}}_j={\bm{u}}$, $D_j({\bm{u}})\to^Pd_j({\bm{u}})$ as $n_j\to\infty,\ j\in\mathcal{J}$ and $J\to\infty$.
	\end{lemma}
	
	\begin{proof}[Proof of Lemma \ref{Lem1}]
		We have
		\begin{equation*}
			E\left[n_{j0}n_j^{-1}P(Y_{ij}>\omega_{(J, n_j)}\mid {\bm{U}}_j={\bm{u}})^{-1}\ \middle|\ {\bm{U}}_j={\bm{u}}\right]=1
		\end{equation*}
		and
		\begin{equation*}
			{\rm{cov}}\left[n_{j0}n_j^{-1}P(Y_{ij}>\omega_{(J, n_j)}\mid {\bm{U}}_j={\bm{u}})^{-1}\ \middle|\ {\bm{U}}_j={\bm{u}}\right]=n_j^{-1}P(Y_{ij}>\omega_{(J, n_j)}\mid {\bm{U}}_j={\bm{u}})^{-1}-n_j^{-1},
		\end{equation*}
		which converges to 0 as $J\to\infty$ and $n_j\to\infty$ from (A3). Accordingly, under given ${\bm{U}}_j={\bm{u}}$, as $J\to\infty$ and $n_j\to\infty$,
		\begin{equation*}
			n_{j0}n_j^{-1}P(Y_{ij}>\omega_{(J, n_j)}\mid {\bm{U}}_j={\bm{u}})^{-1}\xrightarrow{P}1.
		\end{equation*}
		Thus, Lemma \ref{Lem1} is proved by combining this result with (A4).
	\end{proof}
	
	For each $j\in\mathcal{J}$, we denote
	\begin{equation}
		H_j({\bm{u}})\coloneqq n_0^{-1}\sum_{i=1}^{n_j}h_j(Y_{ij}, {\bm{u}}, {\bm{X}}_{ij}),\label{SMEq3.0.1}
	\end{equation}
	where 
	\begin{equation*}
		h_j(y, {\bm{u}}, {\bm{x}})\coloneqq\left[\log\gamma({\bm{u}}, {\bm{x}})+\gamma({\bm{u}}, {\bm{x}})^{-1}\log\frac{y}{\omega_{(J, n_j)}}\right]I(y>\omega_{(J, n_j)}),
	\end{equation*}
	which satisfies
	\begin{equation*}
		\nabla_{\bm{u}}h_j(y, {\bm{u}}, {\bm{x}})=\left[1-\gamma({\bm{u}}, {\bm{x}})^{-1}\log\frac{y}{\omega_{(J, n_j)}}\right]I(y>\omega_{(J, n_j)}){\bm{x}}_{\rm{A}}
	\end{equation*}
	and
	\begin{equation*}
		\nabla_{{\bm{u}}{\bm{u}}}^2h_j(y, {\bm{u}}, {\bm{x}})=\gamma({\bm{u}}, {\bm{x}})^{-1}\log\frac{y}{\omega_{(J, n_j)}}I(y>\omega_{(J, n_j)}){\bm{x}}_{\rm{A}}^{\otimes2}.
	\end{equation*}
	In the following Lemmas \ref{Lem2} and \ref{Lem3}, we reveal the asymptotic properties of $H_j,\ j\in\mathcal{J}$.
	
	\begin{lemma}\label{Lem2}
		Suppose that (A1)-(A4) and (A6) hold. Then, under given ${\bm{U}}_j={\bm{u}}$, as $n_j\to\infty,\ j\in\mathcal{J}$ and $J\to\infty$,
		\begin{equation*}
			n_0^{1/2}\nabla H_j({\bm{u}})\xrightarrow{D}N({\bm{0}}, d_j({\bm{u}}){\bm{\Phi}}_{\rm{AA}}({\bm{u}})).
		\end{equation*}
	\end{lemma}
	
	\begin{proof}[Proof of Lemma \ref{Lem2}]
		For each $j\in\mathcal{J}$, $n_0^{1/2}\nabla H_j({\bm{u}})$ can be written as
		\begin{align}
			n_0^{1/2}\nabla H_j({\bm{u}})&=n_0^{-1/2}\sum_{i=1}^{n_j}\nabla_{\bm{u}}h_j(Y_{ij}, {\bm{u}}, {\bm{X}}_{ij})\nonumber\\
			&=D_j(\bm{u})^{1/2} n_j^{-1/2}\sum_{i=1}^{n_j}\frac{\nabla_{\bm{u}}h_j(Y_{ij}, {\bm{u}}, {\bm{X}}_{ij})-E\left[\nabla_{\bm{u}}h_j(Y_{ij}, {\bm{u}}, {\bm{X}}_{ij})\ \middle|\ {\bm{U}}_j={\bm{u}}\right]}{P(Y_{ij}>\omega_{(J, n_j)}\mid {\bm{U}}_j={\bm{u}})^{1/2}}\label{SMEq3.0.2}\\
			&\quad+D_j(\bm{u})^{1/2} \frac{n_j^{1/2}E\left[\nabla_{\bm{u}}h_j(Y_{ij}, {\bm{u}}, {\bm{X}}_{ij})\ \middle|\ {\bm{U}}_j={\bm{u}}\right]}{P(Y_{ij}>\omega_{(J, n_j)}\mid {\bm{U}}_j={\bm{u}})^{1/2}}.\label{SMEq3.0.3}
		\end{align}
		In the following Steps \ref{St2.1} and \ref{St2.2}, we derive the asymptotic distributions of (\ref{SMEq3.0.2}) and (\ref{SMEq3.0.3}) conditional on ${\bm{U}}_j={\bm{u}}$. By combining Lemma \ref{Lem1} and these steps, Lemma \ref{Lem2} holds from Slutsky's theorem.
		
		\begin{step}\label{St2.1}
			For (\ref{SMEq3.0.3}), we show
			\begin{equation*}
				\frac{J^{1/2}n_j^{1/2}E\left[\nabla_{\bm{u}}h_j(Y_{ij}, {\bm{u}}, {\bm{X}}_{ij})\ \middle|\ {\bm{U}}_j={\bm{u}}\right]}{P(Y_{ij}>\omega_{(J, n_j)}\mid {\bm{U}}_j={\bm{u}})^{1/2}}\to{\bm{b}}_{{\rm{A}}j}({\bm{u}})
			\end{equation*}
			as $J\to\infty$ and $n_j\to\infty$. Because ${\bm{U}}_j$ and ${\bm{X}}_{ij}$ are independent, we have
			\begin{align}
				&\frac{J^{1/2}n_j^{1/2}E\left[\nabla_{\bm{u}}h_j(Y_{ij}, {\bm{u}}, {\bm{X}}_{ij})\ \middle|\ {\bm{U}}_j={\bm{u}}\right]}{P(Y_{ij}>\omega_{(J, n_j)}\mid {\bm{U}}_j={\bm{u}})^{1/2}}\nonumber\\
				&\quad=\frac{J^{1/2}n_j^{1/2}E_{{\bm{X}}_{ij}}\left[E\left[\nabla_{\bm{u}}h_j(Y_{ij}, {\bm{u}}, {\bm{X}}_{ij})\ \middle|\ {\bm{U}}_j={\bm{u}}, {\bm{X}}_{ij}\right]\right]}{P(Y_{ij}>\omega_{(J, n_j)}\mid {\bm{U}}_j={\bm{u}})^{1/2}}.\label{SMEq3.0.4}
			\end{align}
			By the integration by parts, we have
			\begin{align*}
				&E\left[\nabla_{\bm{u}}h_j(Y_{ij}, {\bm{u}}, {\bm{x}})\ \middle|\ {\bm{U}}_j={\bm{u}}, {\bm{X}}_{ij}={\bm{x}}\right]\\
				&\quad=\left[\bar{F}(\omega_{(J, n_j)}\mid {\bm{u}}, {\bm{x}})-\gamma({\bm{u}}, {\bm{x}})^{-1}\int_0^\infty\bar{F}(\omega_{(J, n_j)}e^s\mid {\bm{u}}, {\bm{x}})\,ds\right]{\bm{x}}_{\rm{A}},
			\end{align*}
			where $\bar{F}(\cdot\mid {\bm{u}}, {\bm{x}})\coloneqq1-F(\cdot\mid {\bm{u}}, {\bm{x}})$. Furthermore, from (A1), we have
			\begin{align*}
				&\bar{F}(\omega_{(J, n_j)}\mid {\bm{u}}, {\bm{x}})-\gamma({\bm{u}}, {\bm{x}})^{-1}\int_0^\infty\bar{F}(\omega_{(J, n_j)}e^s\mid {\bm{u}}, {\bm{x}})\,ds\\
				&\quad=\left[\frac{c_1({\bm{u}}, {\bm{x}})\gamma({\bm{u}}, {\bm{x}})\beta({\bm{u}}, {\bm{x}})}{1+\gamma({\bm{u}}, {\bm{x}})\beta({\bm{u}}, {\bm{x}})}\omega_{(J, n_j)}^{-1/\gamma({\bm{u}}, {\bm{x}})-\beta({\bm{u}}, {\bm{x}})}\right]\left[1+o(1)\right].
			\end{align*}
			Accordingly, from (A6), (\ref{SMEq3.0.4}) converges to ${\bm{b}}_{{\rm{A}}j}({\bm{u}})$ as $J\to\infty$ and $n_j\to\infty$.
		\end{step}
		
		\begin{step}\label{St2.2}
			For (\ref{SMEq3.0.2}), we show that under given ${\bm{U}}_j={\bm{u}}$, 
			\begin{equation}
				n_j^{-1/2}\sum_{i=1}^{n_j}\frac{\nabla_{\bm{u}}h_j(Y_{ij}, {\bm{u}}, {\bm{X}}_{ij})-E\left[\nabla_{\bm{u}}h_j(Y_{ij}, {\bm{u}}, {\bm{X}}_{ij})\ \middle|\ {\bm{U}}_j={\bm{u}}\right]}{P(Y_{ij}>\omega_{(J, n_j)}\mid {\bm{U}}_j={\bm{u}})^{1/2}}\xrightarrow{D}N({\bm{0}}, {\bm{\Phi}}_{\rm{AA}}({\bm{u}}))\label{SMEq3.0.5}
			\end{equation}
			as $J\to\infty$ and $n_j\to\infty$. Because (\ref{SMEq3.0.5}) is the sum of conditionally independent and identically distributed random vectors, we can apply the central limit theorem. Obviously, the conditional expectation of (\ref{SMEq3.0.5}) is ${\bm{0}}$. Moreover, we obtain
			\begin{align}
				&{\rm{cov}}\left[\frac{\nabla_{\bm{u}}h_j(Y_{ij}, {\bm{u}}, {\bm{X}}_{ij})}{P(Y_{ij}>\omega_{(J, n_j)}\mid {\bm{U}}_j={\bm{u}})^{1/2}}\ \middle|\ {\bm{U}}_j={\bm{u}}\right]\nonumber\\
				&\quad=E\left[\frac{\nabla_{\bm{u}}h_j(Y_{ij}, {\bm{u}}, {\bm{X}}_{ij})^{\otimes2}}{P(Y_{ij}>\omega_{(J, n_j)}\mid {\bm{U}}_j={\bm{u}})}\ \middle|\ {\bm{U}}_j={\bm{u}}\right]\label{SMEq3.0.6}\\
				&\quad\quad-E\left[\frac{\nabla_{\bm{u}}h_j(Y_{ij}, {\bm{u}}, {\bm{X}}_{ij})}{P(Y_{ij}>\omega_{(J, n_j)}\mid {\bm{U}}_j={\bm{u}})^{1/2}}\ \middle|\ {\bm{U}}_j={\bm{u}}\right]^{\otimes2}.\label{SMEq3.0.7}
			\end{align}
			From Step \ref{St2.1}, (\ref{SMEq3.0.7}) converges to ${\bm{O}}$ as $J\to\infty$ and $n_j\to\infty$. Therefore, we show that (\ref{SMEq3.0.6}) converges to ${\bm{\Phi}}_{\rm{AA}}({\bm{u}})$ as $J\to\infty$ and $n_j\to\infty$. From Eq. (9) of our main article, we have
			\begin{equation}
				{\bm{\xi}}_j({\bm{u}},{\bm{x}})\coloneqq E\left[\nabla_{\bm{u}}h_j(Y_{ij}, {\bm{u}}, {\bm{X}}_{ij})^{\otimes2}\ \middle|\ {\bm{U}}_j={\bm{u}}, {\bm{X}}_{ij}={\bm{x}}, Y_{ij}>\omega_{(J, n_j)}\right]\to{\bm{x}}_{\rm{A}}^{\otimes2}\label{SMEq3.0.8}
			\end{equation}
			uniformly for all ${\bm{x}}\in\mathbb{R}^p$ as $J\to\infty$ and $n_j\to\infty$. In addition, from (A2), we have
			\begin{equation}
				\delta_j({\bm{u}},{\bm{x}})\coloneqq\frac{P(Y_{ij}>\omega_{(J, n_j)}\mid {\bm{U}}_j={\bm{u}}, {\bm{X}}_{ij}={\bm{x}})}{P(Y_{ij}>\omega_{(J, n_j)}\mid {\bm{U}}_j={\bm{u}})}\to\delta({\bm{u}}, {\bm{x}})\label{SMEq3.0.9}
			\end{equation}
			uniformly for all ${\bm{x}}\in\mathbb{R}^p$ as $J\to\infty$ and $n_j\to\infty$. Now, (\ref{SMEq3.0.6}) can be written as
			\begin{equation*}
				E\left[\frac{\nabla_{\bm{u}}h_j(Y_{ij}, {\bm{u}}, {\bm{X}}_{ij})^{\otimes2}}{P(Y_{ij}>\omega_{(J, n_j)}\mid {\bm{U}}_j={\bm{u}})}\ \middle|\ {\bm{U}}_j={\bm{u}}\right]=E_{{\bm{X}}_{ij}}\left[\delta_j({\bm{u}}, {\bm{X}}_{ij}){\bm{\xi}}_j({\bm{u}}, {\bm{X}}_{ij})\right].
			\end{equation*}
			From (\ref{SMEq3.0.8}) and (\ref{SMEq3.0.9}), (\ref{SMEq3.0.6}) then converges to ${\bm{\Phi}}_{\rm{AA}}({\bm{u}})$ as $J\to\infty$ and $n_j\to\infty$.
		\end{step}
	\end{proof}
	
	\begin{lemma}\label{Lem3}
		Suppose that (A1)-(A4) hold. Then, under given ${\bm{U}}_j={\bm{u}}$, as $n_j\to\infty,\ j\in\mathcal{J}$ and $J\to\infty$,
		\begin{equation*}
			\nabla^2H_j({\bm{u}})\xrightarrow{P}d_j({\bm{u}}){\bm{\Phi}}_{\rm{AA}}({\bm{u}}).
		\end{equation*}
	\end{lemma}
	
	\begin{proof}[Proof of Lemma \ref{Lem3}]
		For each $j\in\mathcal{J}$, $\nabla^2H_j({\bm{u}})$ can be written as
		\begin{align*}
			\nabla^2H_j({\bm{u}})&=n_0^{-1}\sum_{i=1}^{n_j}\nabla_{\bm{uu}}^2h_j(Y_{ij}, {\bm{u}}, {\bm{X}}_{ij})\\
			&=D_j(\bm{u})n_j^{-1}\sum_{i=1}^{n_j}\frac{\nabla_{\bm{uu}}^2h_j(Y_{ij}, {\bm{u}}, {\bm{X}}_{ij})}{P(Y_{ij}>\omega_{(J, n_j)}\mid {\bm{U}}_j={\bm{u}})}.
		\end{align*}
		We show that under given ${\bm{U}}_j={\bm{u}}$,
		\begin{equation}
			n_j^{-1}\sum_{i=1}^{n_j}\frac{\nabla_{\bm{uu}}^2h_j(Y_{ij}, {\bm{u}}, {\bm{X}}_{ij})}{P(Y_{ij}>\omega_{(J, n_j)}\mid {\bm{U}}_j={\bm{u}})}\xrightarrow{P}{\bm{\Phi}}_{\rm{AA}}({\bm{u}})\label{SMEq3.0.10}
		\end{equation}
		as $J\to\infty$ and $n_j\to\infty$. From Eq. (9) of our main article, we have
		\begin{equation*}
			{\bm{\xi}}_j^{(1)}({\bm{u}}, {\bm{x}})\coloneqq E\left[\nabla_{\bm{uu}}^2h_j(Y_{ij}, {\bm{u}}, {\bm{x}})\ \middle|\ {\bm{U}}_j={\bm{u}}, {\bm{X}}_{ij}={\bm{x}}, Y_{ij}>\omega_{(J, n_j)}\right]\to{\bm{x}}_{\rm{A}}^{\otimes2}
		\end{equation*}
		and
		\begin{equation*}
			{\bm{\xi}}_j^{(2)}({\bm{u}}, {\bm{x}})\coloneqq E\left[{\rm{vec}}\left[\nabla_{\bm{uu}}^2h_j(Y_{ij}, {\bm{u}}, {\bm{x}})\right]^{\otimes2}\ \middle|\ {\bm{U}}_j={\bm{u}}, {\bm{X}}_{ij}={\bm{x}}, Y_{ij}>\omega_{(J, n_j)}\right]\to2{\rm{vec}}\left({\bm{x}}_{\rm{A}}^{\otimes2}\right)^{\otimes2}
		\end{equation*}
		uniformly for all ${\bm{x}}\in\mathbb{R}^p$ as $J\to\infty$ and $n_j\to\infty$. Now, (\ref{SMEq3.0.10}) has the form of the sum of conditionally independent and identically distributed random vectors, and ${\bm{U}}_j$ and ${\bm{X}}_{ij}$ are independent. These facts yield that for (\ref{SMEq3.0.10}),
		\begin{align*}
			E\left[n_j^{-1}\sum_{i=1}^{n_j}\frac{\nabla_{\bm{uu}}^2h_j(Y_{ij}, {\bm{u}}, {\bm{X}}_{ij})}{P(Y_{ij}>\omega_{(J, n_j)}\mid {\bm{U}}_j={\bm{u}})}\ \middle|\ {\bm{U}}_j={\bm{u}}\right]&=E\left[\delta_j({\bm{u}}, {\bm{X}}_{ij}){\bm{\xi}}_j^{(1)}({\bm{u}}, {\bm{X}}_{ij})\right]\\
			&\to{\bm{\Phi}}_{\rm{AA}}({\bm{u}})
		\end{align*}
		and
		\begin{align*}
			&{\rm{cov}}\left[n_j^{-1}\sum_{i=1}^{n_j}\frac{{\rm{vec}}\left[\nabla_{\bm{uu}}^2h_j(Y_{ij}, {\bm{u}}, {\bm{X}}_{ij})\right]}{P(Y_{ij}>\omega_{(J, n_j)}\mid {\bm{U}}_j={\bm{u}})}\ \middle|\ {\bm{U}}_j={\bm{u}}\right]\\
			\begin{split}
				&=n_j^{-1}P(Y_{ij}>\omega_{(J, n_j)}\mid{\bm{U}}_j={\bm{u}})^{-1}E\left[\delta_j({\bm{u}}, {\bm{X}}_{ij}){\bm{\xi}}_j^{(2)}({\bm{u}}, {\bm{X}}_{ij})\right]\\
				&\quad-n_j^{-1}E\left[\delta_j({\bm{u}}, {\bm{X}}_{ij}){\rm{vec}}\left[{\bm{\xi}}_j^{(1)}({\bm{u}}, {\bm{X}}_{ij})\right]\right]^{\otimes2}
			\end{split}\\
			&\to{\bm{O}}
		\end{align*}
		as $J\to\infty$ and $n_j\to\infty$, where $\delta_j$ is defined in (\ref{SMEq3.0.9}). Therefore, (\ref{SMEq3.0.10}) holds. By combining Lemma \ref{Lem1} and (\ref{SMEq3.0.10}), we then obtain Lemma \ref{Lem3}.
	\end{proof}
	
	For each $j\in\mathcal{J}$, we denote the minimizer of $H_j({\bm{u}})$ defined in (\ref{SMEq3.0.1}) as $\dot{\bm{U}}_j$, which satisfies the following Lemma \ref{Lem4}.
	
	\begin{lemma}\label{Lem4}
		Suppose that (A1)-(A4) and (A6) hold. Then, as $n_j\to\infty,\ j\in\mathcal{J}$ and $J\to\infty$, $n_0^{1/2}(\dot{\bm{U}}_j-{\bm{U}}_j)=O_P(1)$ uniformly for all $j\in\mathcal{J}$.
	\end{lemma}
	
	\begin{proof}[Proof of Lemma \ref{Lem4}]
		We show that under given ${\bm{U}}_j={\bm{u}}$, $n_0^{1/2}(\dot{\bm{U}}_j-{\bm{u}})=O_P(1)$ uniformly for all $j\in\mathcal{J}$ and ${\bm{u}}\in\mathbb{R}^{p_{\rm{A}}}$. By the Taylor expansion of $H_j({\bm{u}})$, we have
		\begin{equation*}
			H_j(n_0^{-1/2}{\bm{s}}+{\bm{u}})=H_j({\bm{u}})+n_0^{-1}{\bm{s}}^\top\left[n_0^{1/2}\nabla H_j({\bm{u}})\right]+2^{-1}n_0^{-1}{\bm{s}}^\top\nabla^2H_j({\bm{u}}){\bm{s}}+o_P(1)
		\end{equation*}
		for any ${\bm{s}}\in\mathbb{R}^{p_{\rm{A}}}$ and $j\in\mathcal{J}$. From Lemmas \ref{Lem2} and \ref{Lem3}, we have that for any $\ >0$, there exists a large constant $B>0$ such that for any $j\in\mathcal{J}$ and ${\bm{u}}\in\mathbb{R}^{p_{\rm{A}}}$,
		\begin{equation}
			\liminf_{n_j\to\infty,\ j\in\mathcal{J},\ J\to\infty}P\left(\inf_{{\bm{s}}\in\mathbb{R}^{p_{\rm{A}}}: \left\lVert{\bm{s}}\right\rVert=B}H_j(n_0^{-1/2}{\bm{s}}+{\bm{u}})>H_j({\bm{u}})\ \middle|\ {\bm{U}}_j={\bm{u}}\right)\geq1-\varepsilon.\label{SMEq3.0.11}
		\end{equation}
		We assume that for all ${\bm{u}}\in\mathbb{R}^{p_{\rm{A}}}$, $\nabla^2H_j({\bm{u}})$ is the positive definite matrix, which implies that $H_j({\bm{u}})$ is the strictly convex function. Therefore, $\dot{\bm{U}}_j$ is the unique global minimizer of $H_j({\bm{u}})$. Then, we obtain Lemma \ref{Lem4} (see, the proof of Theorem 1 of Fan and Li 2001).
	\end{proof}
	
	To show Lemma \ref{Lem6} below, we use the result of the following Laplace approximation. The proof is described in (2.6) of Tierney, Kass, and Kadane (1989) and Appendix A of Miyata (2004).
	
	\begin{lemma}\label{Lem5}
		For any smooth functions $g$, $c$ and $h: \mathbb{R}^d\to\mathbb{R}$,
		\begin{align*}
			&\frac{\int g({\bm{u}})c({\bm{u}})\exp\left[-nh({\bm{u}})\right]d{\bm{u}}}{\int c({\bm{u}})\exp\left[-nh({\bm{u}})\right]d{\bm{u}}}\\
			\begin{split}
				&=g(\dot{\bm{u}})+\frac{\nabla g(\dot{\bm{u}})^\top\nabla^2h(\dot{\bm{u}})^{-1}\nabla c(\dot{\bm{u}})}{nc(\dot{\bm{u}})}\\
				&\quad+\frac{{\rm{tr}}\left[\nabla^2h(\dot{\bm{u}})^{-1}\nabla^2g(\dot{\bm{u}})\right]}{2n}-\frac{\nabla g(\dot{\bm{u}})^\top\nabla^2h(\dot{\bm{u}})^{-1}{\bm{a}}(\dot{\bm{u}})}{2n}+O(n^{-2}),
			\end{split}
		\end{align*}
		where $\dot{\bm{u}}$ is the minimizer of $h({\bm{u}})$, ${\bm{a}}(\dot{\bm{u}})$ is the $d\times1$ vector with the $k$th entry equal to ${\rm{tr}}[\nabla^2h(\dot{\bm{u}})^{-1}\nabla^3h(\dot{\bm{u}})_{[k]}]$, $\nabla^3 h(\dot{\bm{u}})_{[k]}$ is the $d\times d$ matrix with the $(i, j)$ entry equal to the $(i, j, k)$ entry of $\nabla^3h(\dot{\bm{u}})$, and $\nabla^3h({\bm{u}})$ denotes the $d\times d\times d$ array with the $(i, j, k)$ entry $(\partial^3/\partial u_i\partial u_j\partial u_k)h({\bm{u}})$.
	\end{lemma}
	
	\begin{lemma}\label{Lem6}
		Suppose that (A1)-(A4) and (A6) hold. Then, as $n_j\to\infty,\ j\in\mathcal{J}$ and $J\to\infty$,
		\begin{align}
			&\nabla_{\rm{A}}\ell_j({\bm{\theta}}^0)={\bm{\Sigma}}_0^{-1}\left[{\bm{U}}_j+n_0^{-1}d_j({\bm{U}}_j)^{-1}{\bm{\Phi}}_{\rm{AA}}({\bm{U}}_j)^{-1}{\bm{g}}_{{\rm{A}}j}({\bm{U}}_j)\right]+O_P(n_0^{-1}),\label{SMEq3.0.12}\\
			&\nabla_{\rm{B}}\ell_j({\bm{\theta}}^0)={\bm{g}}_{{\rm{B}}j}({\bm{U}}_j)-{\bm{\Phi}}_{\rm{AB}}({\bm{U}}_j)^\top{\bm{\Phi}}_{\rm{AA}}({\bm{U}}_j)^{-1}{\bm{g}}_{{\rm{A}}j}({\bm{U}}_j)+O_P(1)\label{SMEq3.0.13}
		\end{align}
		and
		\begin{align}
			\begin{split}
				\nabla_{\rm{C}}\ell_j({\bm{\theta}}^0)&=2^{-1}{\bm{M}}_*\left({\bm{\Sigma}}_0\otimes{\bm{\Sigma}}_0\right)^{-1}{\rm{vec}}\left[{\bm{U}}_j^{\otimes2}-{\bm{\Sigma}}_0+n_0^{-1}d_j({\bm{U}}_j)^{-1}{\bm{U}}_j{\bm{g}}_{{\rm{A}}j}({\bm{U}}_j)^\top{\bm{\Phi}}_{\rm{AA}}({\bm{U}}_j)^{-1}\right.\\
				&\quad+\left.n_0^{-1}d_j({\bm{U}}_j)^{-1}{\bm{\Phi}}_{\rm{AA}}({\bm{U}}_j)^{-1}{\bm{g}}_{{\rm{A}}j}({\bm{U}}_j){\bm{U}}_j^\top\right]+O_P(n_0^{-1}),\label{SMEq3.0.14}
			\end{split}
		\end{align}
		where
		\begin{equation*}
			{\bm{g}}_{{\rm{K}}j}({\bm{u}})\coloneqq\sum_{i=1}^{n_j}\left[\gamma({\bm{u}}, {\bm{X}}_{ij})^{-1}\log\frac{Y_{ij}}{\omega_{(J, n_j)}}-1\right]I(Y_{ij}>\omega_{(J, n_j)}){\bm{X}}_{{\rm{K}}ij}\\
		\end{equation*}
		for $j\in\mathcal{J}$ and ${\rm{K}}\in\{{\rm{A}}, {\rm{B}}\}$.
	\end{lemma}
	
	\begin{proof}[Proof of Lemma \ref{Lem6}]
		For each $j\in\mathcal{J}$, $\nabla_{\rm{K}}\ell_j({\bm{\theta}}^0),\ {\rm{K}}\in\{{\rm{A}}, {\rm{B}}, {\rm{C}}\}$ can be written as
		\begin{align*}
			&\nabla_{{\rm{A}}}\ell_j({\bm{\theta}}^0)=\frac{\int_{\mathbb{R}^{p_{\rm{A}}}}{\bm{g}}_{{\rm{A}}j}({\bm{u}})c_D({\bm{u}})\exp\left[-n_0H_j({\bm{u}})\right]d{\bm{u}}}{\int_{\mathbb{R}^{p_{\rm{A}}}}c_D({\bm{u}})\exp\left[-n_0H_j({\bm{u}})\right]d{\bm{u}}},\\
			&\nabla_{{\rm{B}}}\ell_j({\bm{\theta}}^0)=\frac{\int_{\mathbb{R}^{p_{\rm{A}}}}{\bm{g}}_{{\rm{B}}j}({\bm{u}})c_D({\bm{u}})\exp\left[-n_0H_j({\bm{u}})\right]d{\bm{u}}}{\int_{\mathbb{R}^{p_{\rm{A}}}}c_D({\bm{u}})\exp\left[-n_0H_j({\bm{u}})\right]d{\bm{u}}}
		\end{align*}
		and
		\begin{equation*}
			\nabla_{\rm{C}}\ell_j({\bm{\theta}}^0)=\frac{\int_{\mathbb{R}^{p_{\rm{A}}}}{\bm{g}}_{{\rm{C}}j}({\bm{u}})c_D({\bm{u}})\exp\left[-n_0H_j({\bm{u}})\right]d{\bm{u}}}{\int_{\mathbb{R}^{p_{\rm{A}}}}c_D({\bm{u}})\exp\left[-n_0H_j({\bm{u}})\right]d{\bm{u}}}-2^{-1}{\bm{M}}_*{\rm{vec}}\left({\bm{\Sigma}}_0^{-1}\right),
		\end{equation*}
		where $c_D({\bm{u}})\coloneqq\exp(-2^{-1}{\bm{u}}^\top{\bm{\Sigma}}_0^{-1}{\bm{u}})$ and ${\bm{g}}_{{\rm{C}}j}({\bm{u}})\coloneqq2^{-1}{\bm{M}}_*({\bm{\Sigma}}_0\otimes{\bm{\Sigma}}_0)^{-1}{\rm{vec}}({\bm{u}}^{\otimes2})$. For each $j\in\mathcal{J}$, ${\rm{K}}\in\{{\rm{A}}, {\rm{B}}, {\rm{C}}\}$ and $l\in\{1, 2, \ldots, p_{\rm{K}}\}$, we denote the $l$th component of ${\bm{g}}_{{\rm{K}}j}({\bm{u}})$ as $g_{{\rm{K}}j}^{(l)}({\bm{u}})$. From Lemma \ref{Lem5}, we have
		\begin{align}
			&\frac{\int_{\mathbb{R}^{p_{\rm{A}}}}{\bm{g}}_{{\rm{K}}j}({\bm{u}})c_D({\bm{u}})\exp\left[-n_0H_j({\bm{u}})\right]d{\bm{u}}}{\int_{\mathbb{R}^{p_{\rm{A}}}}c_D({\bm{u}})\exp\left[-n_0H_j({\bm{u}})\right]d{\bm{u}}}\nonumber\\
			\begin{split}
				&={\bm{g}}_{{\rm{K}}j}(\dot{\bm{U}}_j)+\left[\frac{\nabla g_{{\rm{K}}j}^{(l)}(\dot{\bm{U}}_j)^\top\nabla^2H_j(\dot{\bm{U}}_j)^{-1}\nabla c_D(\dot{\bm{U}}_j)}{n_0c_D(\dot{\bm{U}}_j)}\right]_{p_{\rm{K}}\times1,\ 1\leq l\leq p_{\rm{K}}}\\
				&\quad+\left[\frac{{\rm{tr}}\left[\nabla^2H_j(\dot{\bm{U}}_j)^{-1}\nabla^2g_{{\rm{K}}j}^{(l)}(\dot{\bm{U}}_j)\right]}{2n_0}\right]_{p_{\rm{K}}\times1,\ 1\leq l\leq p_{\rm{K}}}\\
				&\quad-\left[\frac{\nabla g_{{\rm{K}}j}^{(l)}(\dot{\bm{U}}_j)^\top\nabla^2H_j(\dot{\bm{U}}_j)^{-1}{\bm{a}}_j(\dot{\bm{U}}_j)}{2n_0}\right]_{p_{\rm{K}}\times1,\ 1\leq l\leq p_{\rm{K}}}+O(n_0^{-2})\label{SMEq3.0.15}
			\end{split}
		\end{align}
		for $j\in\mathcal{J}$ and ${\rm{K}}\in\{{\rm{A}}, {\rm{B}}, {\rm{C}}\}$. In the following Steps \ref{St6.1}-\ref{St6.3}, we evaluate $\nabla_{\rm{K}}\ell_j({\bm{\theta}}^0)$ based on (\ref{SMEq3.0.15}).
		
		\setcounter{step}{0}
		\begin{step}\label{St6.1}
			We apply (\ref{SMEq3.0.15}) to $\nabla_{\rm{A}}\ell_j({\bm{\theta}}^0)$. By straightforward calculation, we have ${\bm{g}}_{{\rm{A}}j}(\dot{\bm{U}}_j)={\bm{0}}$, 
			\begin{equation*}
				\left[\frac{\nabla g_{{\rm{A}}j}^{(l)}(\dot{\bm{U}}_j)^\top\nabla^2H_j(\dot{\bm{U}}_j)^{-1}\nabla c_D(\dot{\bm{U}}_j)}{n_0c_D(\dot{\bm{U}}_j)}\right]_{p_{\rm{A}}\times1,\ 1\leq l\leq p_{\rm{A}}}={\bm{\Sigma}}_0^{-1}\dot{\bm{U}}_j
			\end{equation*}
			and 
			\begin{equation*}
				\frac{{\rm{tr}}\left[\nabla^2H_j(\dot{\bm{U}}_j)^{-1}\nabla^2g_{{\rm{A}}j}^{(l)}(\dot{\bm{U}}_j)\right]}{2n_0}-\frac{\nabla g_{{\rm{A}}j}^{(l)}(\dot{\bm{U}}_j)^\top\nabla^2H_j(\dot{\bm{U}}_j)^{-1}{\bm{a}}_j(\dot{\bm{U}}_j)}{2n_0}=0
			\end{equation*}
			for $l\in\{1, 2, \ldots, p_{\rm{A}}\}$. Furthermore, from Lemmas \ref{Lem3} and \ref{Lem4}, by the Taylor expansion of $\nabla H_j({\bm{u}})$, we obtain
			\begin{equation}
				\dot{\bm{U}}_j={\bm{U}}_j+n_0^{-1}d_j({\bm{U}}_j)^{-1}{\bm{\Phi}}_{\rm{AA}}({\bm{U}}_j)^{-1}{\bm{g}}_{{\rm{A}}j}({\bm{U}}_j)+O_P(n_0^{-1}).\label{SMEq3.0.16}
			\end{equation}
			Consequently, we obtain (\ref{SMEq3.0.12}).
		\end{step}
		
		\begin{step}\label{St6.2}
			In this step, (\ref{SMEq3.0.15}) is applied to $\nabla_{\rm{B}}\ell_j({\bm{\theta}}^0)$. From Lemma \ref{Lem4}, (\ref{SMEq3.0.16}) and the similar results to Lemma \ref{Lem3}, by the Taylor expansion of ${\bm{g}}_{{\rm{B}}j}({\bm{u}})$, we obtain
			\begin{align*}
				{\bm{g}}_{{\rm{B}}j}(\dot{\bm{U}}_j)&={\bm{g}}_{{\rm{B}}j}({\bm{U}}_j)-n_0d_j({\bm{U}}_j){\bm{\Phi}}_{\rm{AB}}({\bm{U}}_j)^\top\left(\dot{\bm{U}}_j-{\bm{U}}_j\right)\left[1+o_P(1)\right]\\
				&={\bm{g}}_{{\rm{B}}j}({\bm{U}}_j)-{\bm{\Phi}}_{\rm{AB}}({\bm{U}}_j)^\top{\bm{\Phi}}_{\rm{AA}}({\bm{U}}_j)^{-1}{\bm{g}}_{{\rm{A}}j}({\bm{U}}_j)+O_P(1).
			\end{align*}
			In addition, we have
			\begin{align*}
				\begin{split}
					&\frac{\nabla g_{{\rm{B}}j}^{(l)}(\dot{\bm{U}}_j)^\top\nabla^2H_j(\dot{\bm{U}}_j)^{-1}\nabla c_D(\dot{\bm{U}}_j)}{n_0c_D(\dot{\bm{U}}_j)}+\frac{{\rm{tr}}\left[\nabla^2H_j(\dot{\bm{U}}_j)^{-1}\nabla^2g_{{\rm{B}}j}^{(l)}(\dot{\bm{U}}_j)\right]}{2n_0}\\
					&\quad-\frac{\nabla g_{{\rm{B}}j}^{(l)}(\dot{\bm{U}}_j)^\top\nabla^2H_j(\dot{\bm{U}}_j)^{-1}{\bm{a}}_j(\dot{\bm{U}}_j)}{2n_0}=O_P(1)
				\end{split}
			\end{align*}
			for $l\in\{1, 2, \ldots, p_{\rm{B}}\}$. Therefore, (\ref{SMEq3.0.13}) is obtained.
		\end{step}
		
		\begin{step}\label{St6.3}
			In the last step, we calculate $\nabla_{\rm{C}}\ell_j({\bm{\theta}}^0)$ according to (\ref{SMEq3.0.15}). From (\ref{SMEq3.0.16}), we have
			\begin{align*}
				{\bm{g}}_{{\rm{C}}j}(\dot{\bm{U}}_j)&=2^{-1}{\bm{M}}_*\left({\bm{\Sigma}}_0\otimes{\bm{\Sigma}}_0\right)^{-1}{\rm{vec}}\left(\dot{\bm{U}}_j^{\otimes2}\right)\\
				\begin{split}
					&=2^{-1}{\bm{M}}_*\left({\bm{\Sigma}}_0\otimes{\bm{\Sigma}}_0\right)^{-1}\\
					&\quad\times{\rm{vec}}\left[{\bm{U}}_j^{\otimes2}+n_0^{-1}d_j({\bm{U}}_j)^{-1}{\bm{U}}_j{\bm{g}}_{{\rm{A}}j}({\bm{U}}_j)^\top{\bm{\Phi}}_{\rm{AA}}({\bm{U}}_j)^{-1}\right.\\
					&\quad\left.+n_0^{-1}d_j({\bm{U}}_j)^{-1}{\bm{\Phi}}_{\rm{AA}}({\bm{U}}_j)^{-1}{\bm{g}}_{{\rm{A}}j}({\bm{U}}_j){\bm{U}}_j^\top\right]+O_P(n_0^{-1}).
				\end{split}
			\end{align*}
			Additionally, we have
			\begin{align*}
				\begin{split}
					&\frac{\nabla g_{{\rm{C}}j}^{(l)}(\dot{\bm{U}}_j)^\top\nabla^2H_j(\dot{\bm{U}}_j)^{-1}\nabla c_D(\dot{\bm{U}}_j)}{n_0c_D(\dot{\bm{U}}_j)}+\frac{{\rm{tr}}\left[\nabla^2H_j(\dot{\bm{U}}_j)^{-1}\nabla^2g_{{\rm{C}}j}^{(l)}(\dot{\bm{U}}_j)\right]}{2n_0}\\
					&\quad-\frac{\nabla g_{{\rm{C}}j}^{(l)}(\dot{\bm{U}}_j)^\top\nabla^2H_j(\dot{\bm{U}}_j)^{-1}{\bm{a}}_j(\dot{\bm{U}}_j)}{2n_0}=O_P(n_0^{-1})
				\end{split}
			\end{align*}
			for $l\in\{1, 2, \ldots, p_{\rm{C}}\}$. Thus, (\ref{SMEq3.0.14}) is shown.
		\end{step}
	\end{proof}
	
	Lemma \ref{Lem6} leads to the following Lemmas \ref{Lem7}-\ref{Lem9}.
	
	\begin{lemma}\label{Lem7}
		Suppose that (A1)-(A6) hold. Then, as $n_j\to\infty,\ j\in\mathcal{J}$ and $J\to\infty$,
		\begin{equation*}
			J^{-1/2}\nabla_{\rm{A}}\ell({\bm{\theta}}^0)+n_0^{-1/2}{\bm{\Delta}}_{\rm{A}}^{-1}{\bm{b}}_{\rm{A}}\xrightarrow{D}N({\bm{0}}, {\bm{\Delta}}_{\rm{A}}^{-1}).
		\end{equation*}
	\end{lemma}
	
	\begin{proof}[Proof of Lemma \ref{Lem7}]
		Let denote
		\begin{equation*}
			{\bm{Z}}\coloneqq J^{-1/2}\sum_{j=1}^Jn_0^{-1/2}d_j({\bm{U}}_j)^{-1}{\bm{\Phi}}_{\rm{AA}}({\bm{U}}_j)^{-1}{\bm{g}}_{{\rm{A}}j}({\bm{U}}_j).
		\end{equation*}
		From Lemma \ref{Lem6}, we then have
		\begin{equation}
			J^{-1/2}\nabla_{\rm{A}}\ell({\bm{\theta}}^0)=J^{-1/2}\sum_{j=1}^J{\bm{\Sigma}}_0^{-1}{\bm{U}}_j+n_0^{-1/2}{\bm{\Sigma}}_0^{-1}{\bm{Z}}\left[1+o_P(1)\right].\label{SMEq3.0.17}
		\end{equation}
		From the reproductive property of the normal distribution, the first term on the right-hand side of (\ref{SMEq3.0.17}) converges to $N({\bm{0}}, {\bm{\Sigma}}_0^{-1})$ in distribution as $J\to\infty$. From the proof of Lemma \ref{Lem2}, for the second term of the right-hand side of (\ref{SMEq3.0.17}), we have
		\begin{align*}
			E\left[{\bm{Z}}\right]&=J^{-1}\sum_{j=1}^JE\left[d_j({\bm{U}}_j)^{-1}{\bm{\Phi}}_{\rm{AA}}({\bm{U}}_j)^{-1}E\left[J^{1/2}n_0^{-1/2}{\bm{g}}_{{\rm{A}}j}({\bm{U}}_j)\ \middle|\ {\bm{U}}_j\right]\right]\\
			&=-\left(J^{-1}\sum_{j=1}^JE\left[d_j({\bm{U}}_j)^{-1/2}{\bm{\Phi}}_{\rm{AA}}({\bm{U}}_j)^{-1}{\bm{b}}_{{\rm{A}}j}({\bm{U}}_j)\right]\right)\left[1+o(1)\right]\\
			&=-{\bm{b}}_{\rm{A}}\left[1+o(1)\right]
		\end{align*}
		and ${\rm{cov}}[n_0^{-1/2}{\bm{Z}}]\to{\bm{O}}$, which implies that the sum of $n_0^{-1/2}{\bm{\Sigma}}_0^{-1}{\bm{b}}_{\rm{A}}$ and the second term on the right-hand side of (\ref{SMEq3.0.17}) converges to ${\bm{0}}$ in probability as $n_j\to\infty,\ j\in\mathcal{J}$ and $J\to\infty$. Thus, Lemma {\ref{Lem7}} is shown.
	\end{proof}
	
	\begin{lemma}\label{Lem8}
		Suppose that (A1)-(A6) hold. Then, as $n_j\to\infty,\ j\in\mathcal{J}$ and $J\to\infty$,
		\begin{equation*}
			J^{-1/2}n_0^{-1/2}\nabla_{\rm{B}}\ell({\bm{\theta}}^0)\xrightarrow{D}N(-{\bm{\Delta}}_{\rm{B}}^{-1}{\bm{b}}_{\rm{B}}, {\bm{\Delta}}_{\rm{B}}^{-1}).
		\end{equation*}
	\end{lemma}
	
	\begin{proof}[Proof of Lemma \ref{Lem8}]
		We denote 
		\begin{equation*}
			{\bm{W}}_j({\bm{u}})\coloneqq{\bm{g}}_{{\rm{B}}j}({\bm{u}})-{\bm{\Phi}}_{\rm{AB}}({\bm{u}})^\top{\bm{\Phi}}_{\rm{AA}}({\bm{u}})^{-1}{\bm{g}}_{{\rm{A}}j}({\bm{u}}).
		\end{equation*}
		From Lemma \ref{Lem6}, we have
		\begin{equation}
			J^{-1/2}n_0^{-1/2}\nabla_{\rm{B}}\ell({\bm{\theta}}^0)=\left[J^{-1/2}\sum_{j=1}^Jn_0^{-1/2}{\bm{W}}_j({\bm{U}}_j)\right]\left[1+o_P(1)\right].\label{SMEq3.0.18}
		\end{equation}
		Now, the right-hand side of (\ref{SMEq3.0.18}) can be written as 
		\begin{align}
			&J^{-1/2}\sum_{j=1}^Jn_0^{-1/2}{\bm{W}}_j({\bm{U}}_j)\nonumber\\
			&=J^{-1/2}\sum_{j=1}^Jn_{j0}^{1/2}n_0^{-1/2}\left[n_{j0}^{-1/2}{\bm{W}}_j({\bm{U}}_j)-E\left[n_{j0}^{-1/2}{\bm{W}}_j({\bm{U}}_j)\ \middle|\ {\bm{U}}_j\right]\right]\label{SMEq3.0.19}\\
			&\quad+J^{-1}\sum_{j=1}^JE\left[J^{1/2}n_0^{-1/2}{\bm{W}}_j({\bm{U}}_j)\ \middle|\ {\bm{U}}_j\right].\label{SMEq3.0.20}
		\end{align}
		Similar to the proof of Lemma \ref{Lem7}, (\ref{SMEq3.0.20}) converges to $-{\bm{\Delta}}_{\rm{B}}^{-1}{\bm{b}}_{\rm{B}}$ in probability as $n_j\to\infty,\ j\in\mathcal{J}$ and $J\to\infty$. Similar to Lemma \ref{Lem2}, for (\ref{SMEq3.0.19}), we have that under given ${\bm{U}}_j={\bm{u}}_j$,
		\begin{align*}
			&n_{j0}^{-1/2}{\bm{W}}_j({\bm{u}}_j)-E\left[n_{j0}^{-1/2}{\bm{W}}_j({\bm{u}}_j)\ \middle|\ {\bm{U}}_j={\bm{u}}_j\right]\\
			&\quad\xrightarrow{D}N({\bm{0}}, {\bm{\Phi}}_{\rm{BB}}({\bm{u}}_j)-{\bm{\Phi}}_{\rm{AB}}({\bm{u}}_j)^\top{\bm{\Phi}}_{\rm{AA}}({\bm{u}}_j)^{-1}{\bm{\Phi}}_{\rm{AB}}({\bm{u}}_j))
		\end{align*}
		as $J\to\infty$ and $n_j\to\infty$. Therefore, (\ref{SMEq3.0.19}) is the weighted sum of independent and asymptotically identically distributed random vectors, which can be applied the weighted central limit theorem (see, Weber 2006). As a result, (\ref{SMEq3.0.19}) converges to $N({\bm{0}}, {\bm{\Delta}}_{\rm{B}}^{-1})$ in distribution as $n_j\to\infty,\ j\in\mathcal{J}$ and $J\to\infty$. Thus, the proof of Lemma \ref{Lem8} is completed.
	\end{proof}
	
	\begin{lemma}\label{Lem9}
		Suppose that (A1)-(A6) hold. Then, as $n_j\to\infty,\ j\in\mathcal{J}$ and $J\to\infty$,
		\begin{equation*}
			J^{-1/2}\nabla_{\rm{C}}\ell({\bm{\theta}}^0)+n_0^{-1/2}{\bm{\Delta}}_{\rm{C}}^{-1}{\bm{b}}_{\rm{C}}\xrightarrow{D}N({\bm{0}}, {\bm{\Delta}}_{\rm{C}}^{-1}).
		\end{equation*}
	\end{lemma}
	
	\begin{proof}[Proof of Lemma \ref{Lem9}]
		Let denote
		\begin{equation*}
			{\bm{V}}_j({\bm{u}})\coloneqq d_j({\bm{u}})^{-1}\left[{\bm{u}}{\bm{g}}_{{\rm{A}}j}({\bm{u}})^\top{\bm{\Phi}}_{\rm{AA}}({\bm{u}})^{-1}+{\bm{\Phi}}_{\rm{AA}}({\bm{u}})^{-1}{\bm{g}}_{{\rm{A}}j}({\bm{u}}){\bm{u}}^\top\right].
		\end{equation*}
		From Lemma \ref{Lem6}, we obtain
		\begin{align}
			&J^{-1/2}\nabla_{\rm{C}}\ell({\bm{\theta}}^0)\nonumber\\
			&=2^{-1}{\bm{M}}_*\left({\bm{\Sigma}}_0\otimes{\bm{\Sigma}}_0\right)^{-1}J^{-1/2}\sum_{j=1}^J{\rm{vec}}\left({\bm{U}}_j^{\otimes2}-{\bm{\Sigma}}_0\right)\label{SMEq3.0.21}\\
			&\quad+2^{-1}{\bm{M}}_*\left({\bm{\Sigma}}_0\otimes{\bm{\Sigma}}_0\right)^{-1}n_0^{-1/2}\left(J^{-1/2}\sum_{j=1}^Jn_0^{-1/2}{\rm{vec}}\left[{\bm{V}}_j({\bm{U}}_j)\right]\right)\left[1+o_P(1)\right].\label{SMEq3.0.22}
		\end{align}
		${\bm{U}}_j^{\otimes2}$ is distributed as a Wishart distribution with $E[{\bm{U}}_j^{\otimes2}]={\bm{\Sigma}}_0$ and ${\rm{cov}}[{\rm{vec}}({\bm{U}}_j^{\otimes2})]=2({\bm{\Sigma}}_0\otimes{\bm{\Sigma}}_0)$. Therefore, by the central limit theorem, (\ref{SMEq3.0.21}) converges to $N({\bm{0}}, {\bm{\Delta}}_{\rm{C}}^{-1})$ in distribution as $J\to\infty$. Moreover, similar to the proof of Lemma \ref{Lem7}, (\ref{SMEq3.0.22}) is asymptotically equivalent to $-n_0^{-1/2}{\bm{\Delta}}_{\rm{C}}^{-1}{\bm{b}}_{\rm{C}}$. Consequently, we obtain Lemma {\ref{Lem9}}.
	\end{proof}
	
	The above Lemmas \ref{Lem7}-\ref{Lem9} are summarized following two propositions.
	
	\begin{proposition}\label{Pro1}
		Suppose that (A1)-(A6) hold. Then, as $n_j\to\infty,\ j\in\mathcal{J}$ and $J\to\infty$,
		\begin{equation*}
			{\bm{\Upsilon}}_{(J, n_0)}^{-1}\nabla\ell({\bm{\theta}}^0)+\begin{bmatrix}
				n_0^{-1/2}{\bm{\Delta}}_{\rm{A}}^{-1}{\bm{b}}_{\rm{A}}\\
				{\bm{\Delta}}_{\rm{B}}^{-1}{\bm{b}}_{\rm{B}}\\
				n_0^{-1/2}{\bm{\Delta}}_{\rm{C}}^{-1}{\bm{b}}_{\rm{C}}\\
			\end{bmatrix}\xrightarrow{D}N\left({\bm{0}}, 
			\begin{bmatrix}
				{\bm{\Delta}}_{\rm{A}}^{-1} & {\bm{O}} & {\bm{O}}\\
				{\bm{O}} & {\bm{\Delta}}_{\rm{B}}^{-1} & {\bm{O}}\\
				{\bm{O}} & {\bm{O}} & {\bm{\Delta}}_{\rm{C}}^{-1}
			\end{bmatrix}
			\right).
		\end{equation*}
	\end{proposition}
	
	\begin{proof}[Proof of Proposition \ref{Pro1}]
		Similar to Lemmas \ref{Lem7}-\ref{Lem9}, from Lemma \ref{Lem6}, we have
		\begin{align*}
			&{\rm{cov}}\left[J^{-1/2}\nabla_{\rm{A}}\ell({\bm{\theta}}^0),\ J^{-1/2}n_0^{-1/2}\nabla_{\rm{B}}\ell({\bm{\theta}}^0)\right]\to{\bm{O}},\\
			&{\rm{cov}}\left[J^{-1/2}\nabla_{\rm{A}}\ell({\bm{\theta}}^0),\ J^{-1/2}\nabla_{\rm{C}}\ell({\bm{\theta}}^0)\right]\to{\bm{O}}
		\end{align*}
		and
		\begin{equation*}
			{\rm{cov}}\left[J^{-1/2}n_0^{-1/2}\nabla_{\rm{B}}\ell({\bm{\theta}}^0),\ J^{-1/2}\nabla_{\rm{C}}\ell({\bm{\theta}}^0)\right]\to{\bm{O}}
		\end{equation*}
		as $n_j\to\infty,\ j\in\mathcal{J}$ and $J\to\infty$. By combining these results and Lemmas \ref{Lem7}-\ref{Lem9}, we obtain Proposition \ref{Pro1}.
	\end{proof}
	
	\begin{proposition}\label{Pro2}
		Suppose that (A1)-(A6) hold. Then, as $n_j\to\infty,\ j\in\mathcal{J}$ and $J\to\infty$,
		\begin{equation*}
			{\bm{\Upsilon}}_{(J, n_0)}^{-1}\nabla^2\ell({\bm{\theta}}^0){\bm{\Upsilon}}_{(J, n_0)}^{-1}\xrightarrow{P}
			\begin{bmatrix}
				-{\bm{\Delta}}_{\rm{A}}^{-1} & {\bm{O}} & {\bm{O}}\\
				{\bm{O}} & -{\bm{\Delta}}_{\rm{B}}^{-1} & {\bm{O}}\\
				{\bm{O}} & {\bm{O}} & -{\bm{\Delta}}_{\rm{C}}^{-1}
			\end{bmatrix}.
		\end{equation*}
	\end{proposition}
	
	\begin{proof}[Proof of Proposition \ref{Pro2}]
		From Lemma 5 of Nie (2007) and Lemma \ref{Lem5}, the covariance matrix of ${\rm{vec}}({\bm{\Upsilon}}_{(J, n_0)}^{-1}\nabla^2\ell({\bm{\theta}}^0){\bm{\Upsilon}}_{(J, n_0)}^{-1})$ converges to ${\bm{O}}$ as $n_j\to\infty,\ j\in\mathcal{J}$ and $J\to\infty$. Now, we have
		\begin{equation}
			E\left[{\bm{\Upsilon}}_{(J, n_0)}^{-1}\nabla^2\ell({\bm{\theta}}^0){\bm{\Upsilon}}_{(J, n_0)}^{-1}\right]=-E\left[{\bm{\Upsilon}}_{(J, n_0)}^{-1}\nabla\ell({\bm{\theta}}^0)^{\otimes2}{\bm{\Upsilon}}_{(J, n_0)}^{-1}\right]\label{SMEq3.0.23}.
		\end{equation}
		By calculating the right-hand side of (\ref{SMEq3.0.23}) using Lemma \ref{Lem6}, Proposition \ref{Pro2} is obtained.
	\end{proof}
	
	\begin{proof}[Proof of Theorem 1 in Section 2.3 of our main article]
		For any ${\bm{\theta}}\in\mathbb{R}^{p_{\rm{A}}+p_{\rm{B}}+p_{\rm{C}}}$, the Taylor expansion of $\ell({\bm{\theta}})$ around ${\rm {\bm\theta}}={\rm {\bm\theta}}^0$ yields that
		\begin{align*}
			&\ell({\bm{\Upsilon}}_{(J, n_0)}^{-1}{\bm{\theta}}+{\bm{\theta}}^0)\\
			&=\ell({\bm{\theta}}^0)+{\bm{\theta}}^\top{\bm{\Upsilon}}_{(J, n_0)}^{-1}\nabla\ell({\bm{\theta}}^0)+2^{-1}{\bm{\theta}}^\top{\bm{\Upsilon}}_{(J, n_0)}^{-1}\nabla^2\ell({\bm{\theta}}^0){\bm{\Upsilon}}_{(J, n_0)}^{-1}{\bm{\theta}}+o_P(1).
		\end{align*}
		From Propositions \ref{Pro1} and \ref{Pro2}, for any $\varepsilon>0$, there exists a large constant $B>0$ such that 
		\begin{equation}
			\liminf_{n_j\to\infty,\ j\in\mathcal{J},\ J\to\infty}P\left(\inf_{{\bm{\theta}}\in\mathbb{R}^{p_{\rm{A}}+p_{\rm{B}}+p_{\rm{C}}}: \left\lVert{\bm{\theta}}\right\rVert=B}-\ell({\bm{\Upsilon}}_{(J, n_0)}^{-1}{\bm{\theta}}+{\bm{\theta}}^0)>-\ell({\bm{\theta}}^0)\right)\geq1-\varepsilon\label{SMEq3.0.24}
		\end{equation}
		as $n_j\to\infty,\ j\in\mathcal{J}$ and $J\to\infty$. We assume that for all ${\bm{\theta}}$, $-\nabla^2\ell({\bm{\theta}})$ is the positive definite matrix, which implies that $-\ell({\bm{\theta}})$ is the strictly convex function. Therefore, $\hat{\bm{\theta}}$ is the unique global maximizer of $\ell({\bm{\theta}})$. Then, (\ref{SMEq3.0.24}) implies ${\bm{\Upsilon}}_{(J, n_0)}(\hat{\bm{\theta}}-{\bm{\theta}}^0)=O_P(1)$ (see, the proof of Theorem 1 of Fan and Li 2001).
		Because $\hat{\bm{\theta}}$ is the global maximizer of $\ell({\bm{\theta}})$, we have $\nabla\ell(\hat{\bm{\theta}})={\bm{0}}$. From the Taylor expansion of $\ell({\bm{\theta}})$, we have
		\begin{align*}
			{\bm{\Upsilon}}_{(J, n_0)}\left(\hat{\bm{\theta}}-{\bm{\theta}}^0\right)&=-\left[{\bm{\Upsilon}}_{(J, n_0)}^{-1}\nabla^2\ell({\bm{\theta}}^0){\bm{\Upsilon}}_{(J, n_0)}^{-1}\right]^{-1}{\bm{\Upsilon}}_{(J, n_0)}^{-1}\nabla\ell({\bm{\theta}}^0)+o_P(1).
		\end{align*}
		Therefore, by applying Propositions \ref{Pro1} and \ref{Pro2}, we obtain Theorem 1 of our main article.
	\end{proof}

\end{document}